\documentclass{aa}  

\usepackage{graphicx}
\usepackage{txfonts}
\usepackage{lipsum}
\usepackage{lscape}             
\usepackage{placeins}           

\usepackage{natbib}
\usepackage[colorlinks,allcolors=blue,bookmarks=false]{hyperref} 
\usepackage{xcolor}
\newcommand{\chgn}[1]{#1}
\usepackage{amsmath}
\begin{document}


   \title{Deriving volume density profiles of filaments\\from observed surface densities}

   \subtitle{}

%

   \author{Alexander Men'shchikov\inst{1}
        \and Guo-Yin Zhang\inst{2}
        }

   \institute{Universit\'{e} Paris-Saclay, Universit\'{e} Paris Cit\'{e}, CEA, CNRS, AIM, 91191, Gif-sur-Yvette, France\\
      \email{alexander.menshchikov@cea.fr}
   \and National Astronomical Observatories, Chinese Academy of Sciences, A20 Datun Road, Chaoyang District, Beijing 100101, China\\
      \email{zgyin@nao.cas.cn}
   }

   \date{Received / Accepted }
   \offprints{Alexander Men'shchikov}
   \titlerunning{Deriving volume density profiles of filaments}
   \authorrunning{A.~Men'shchikov \and G.-Y.~Zhang}

   \abstract{
     Accurate characterization of filamentary structures in star-forming clouds is essential for understanding star formation.
     Traditional methods fit observed surface density profiles $\Sigma(r)$ with slope $\gamma$ and width $H$ using the Plummer
     function, assuming $\beta=\gamma+1$ and $h\approx H$ for the volume density slope and width. These assumptions break down
     for shallow profiles, with the slope and width relations deviating progressively more for compact and extended filaments,
     respectively. We present a new fitting method that explicitly accounts for finite cylindrical geometry and establishes
     self-consistent empirical relationships between the parameters of $\Sigma(r)$ and those of the volume density profile
     $\rho(r)$ with slope $\beta$ and width $h$. The method was validated on model profiles and applied to selected filaments in
     the California molecular cloud. The slope difference $\delta\equiv\beta-\gamma$ falls below unity for shallow ($\beta\lesssim
     2$) and compact profiles; $h$ and $H$ can differ by over an order of magnitude for extended filaments with shallow slopes.
     Accurate parameter recovery requires high resolvedness $\mathcal{R}\equiv H/O\gtrsim 10$ (where $O$ is the beam width); at
     lower resolvedness, slopes are severely overestimated and filaments remain unresolved even when $H\gg O$. The traditional
     Plummer function yields systematically overestimated slopes. Accurate deconvolution requires a priori knowledge of the true
     parameters, creating a fundamental circular problem whose only robust solution is obtaining sufficiently high angular
     resolution. Current far-infrared observations typically lack sufficient resolution, and some previously reported filament
     properties may require reinterpretation.
   }

   \keywords{Stars: formation -- Infrared: ISM -- Submillimeter: ISM -- Methods: data analysis -- Techniques: image processing --
             Techniques: photometric}

   \maketitle


\section{Introduction}
\label{introduction}

Far-infrared observations with the \emph{Herschel} Space Observatory have greatly advanced studies of star formation in our Galaxy
\citep{Pilbratt_etal2010,Andre_etal2010, Molinari_etal2010}. Numerous star-forming regions have been studied in detail
\citep[e.g.,][]{Hennemann_etal2012,Schneider_etal2012,Andre_etal2014,Zhang_etal2024} to understand how both low- and high-mass
stars form in molecular clouds. Most prestellar cores and protostars are found in ubiquitous filamentary structures, making it
essential to determine the physical properties of filaments \citep[e.g.,][]{Men'shchikov_etal2010, Ko"nyves_etal2015,
Zhang_etal2020}. This is typically accomplished by deriving surface density images from multiwavelength \emph{Herschel}
observations, identifying filament crests (skeletons), extracting surface density profiles $\Sigma_{\rm O}(r)$, and fitting the
observed profiles with an analytical model \citep[e.g.,][]{Arzoumanian_etal2011, Palmeirim_etal2013, Arzoumanian_etal2019}.

\begin{figure*}[ht!]
\centerline{\resizebox{0.3355\hsize}{!}{\includegraphics{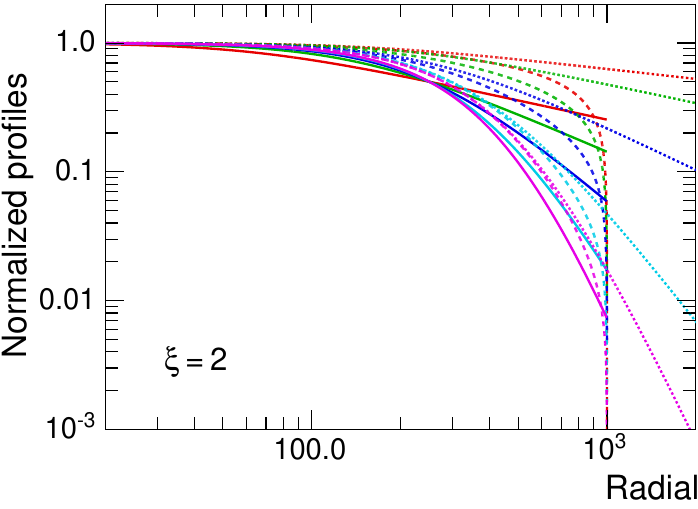}}
            \resizebox{0.3224\hsize}{!}{\includegraphics{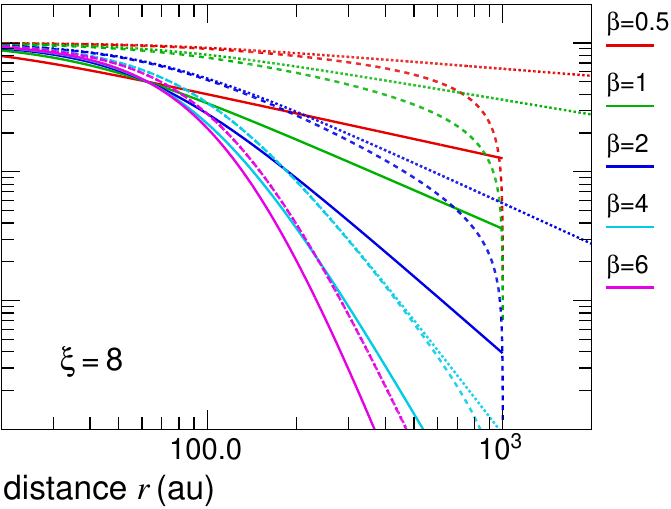}}}
\caption{
Analytical profiles of volume and surface densities (solid and dashed curves, respectively) for structures with a circular boundary
at $R = 10^{3}$ au and representative values of $\xi\equiv R/h$ (defined in Sect.~\ref{fittingfun}) and $\beta$.
Both profile types are computed from the analytical Eqs.~(\ref{volume_density}) and (\ref{surface_density_fittingfun}).
They are shown in the same panels to enable their direct comparison: the slopes and widths of $\rho(r)$ and $\Sigma(r)$
differ systematically, increasingly so toward shallower $\beta$ and larger $\xi$.
For reference, the dotted curves display surface density profiles of
fictitious ``infinite'' density distributions with no outer boundary, corresponding to the case where the square-root term in
Eq.~(\ref{surface_density_fittingfun}) is omitted. The differences between the finite and infinite profiles highlight the
systematic inaccuracies introduced by traditional fitting with Eq.~(\ref{plummerfun}). This effect is most pronounced for
shallow slopes ($\beta \lesssim 4$), where the presence of the outer boundary alters the shapes of the infinite profiles
starting from their deep interiors ($0.2R \lesssim r\rightarrow R$).
} 
\label{vsdensprofiles}
\end{figure*}

Previous studies have used the Plummer function to fit surface density profiles and simultaneously derive volume density profiles:
\begin{equation}
\Sigma(r) = \Sigma_{\rm C}\left(1+\left(r/r_{\rm c}\right)^2\right)^{-(\beta-1)/2},
\label{plummerfun}
\end{equation} 
where $\Sigma_{\rm C}$ is the filament crest surface density, $r$ is the radial distance from the crest, $r_{\rm c}$ is the radius
of the dense core region, and $\beta$ is the power-law slope of the volume density $\rho(r)$, implicitly assumed to be related to
the slope $\gamma$ of the observed $\Sigma_{\rm O}(r)$ as $\gamma = \beta - 1$. However, Eq.~(\ref{plummerfun}) suffers from
several practical limitations when applied to observed finite filaments. First, it does not represent surface density profiles of
finite structures. The function becomes a power law $\Sigma(r) \propto r^{-(\beta-1)}$ at large radii $r \gg r_{\rm c}$. This
unbounded power-law behavior cannot describe filaments treated as finite structures with an outer radius $R$, where
$\Sigma(r)\rightarrow 0$ as $r\rightarrow R$. It is well known that surface densities of finite structures deviate steeply from
power laws as $r \rightarrow R$ (where $\Sigma(r) \rightarrow 0$), even when their volume densities $\rho(r)$ follow a pure power
law truncated at the boundary (Fig.~\ref{vsdensprofiles}). Second, the assumption $\gamma = \beta - 1$ incorporated in
Eq.~(\ref{plummerfun}) is problematic because for $\beta < 1$ it implies $\gamma < 0$ and an unphysical rising $\Sigma(r)$ with
distance. Third, previous studies assumed that the physical widths of the $\rho(r)$ distributions are identical to those of the
$\Sigma(r)$ profiles, which is approximately valid only for $\beta \gtrsim 2$ (Fig.~\ref{vsdensprofiles}). These contradictory and
incorrect assumptions have likely introduced systematic errors and biases in derived filament parameters, affecting estimates of
their fundamental physical properties.

This paper presents a new fitting method that accurately describes surface density profiles of finite structures and ensures full
consistency between $\Sigma(r)$ and $\rho(r)$, thereby enabling reliable derivation of physical volume density profiles from
observed surface densities. We describe the new fitting function and strategy in Sect.~\ref{method}, present validation of the
method using profiles of model filaments in Sect.~\ref{validation}, discuss angular resolution effects and problems with
traditional fitting and deconvolution approaches in Sect.~\ref{discussion}, and summarize our conclusions in
Sect.~\ref{conclusions}. We also derive a deconvolution formula for filament widths in Appendix~\ref{deconvolution}.


\section{Method}
\label{method}

\subsection{Fitting function}
\label{fittingfun}

Assuming that the volume density distribution $\rho(r)$ of a filament has a nonuniform cylindrical geometry (generally, a curved
cylinder with varying circular cross-sections) truncated at radius $R$, the surface density profile $\Sigma(r)$ is obtained by
integrating the densities along the line of sight within the boundary,
\begin{equation}
\Sigma(r) = 2\int_{0}^{L(r)\!}\rho(r^2 + x^2)^{1/2}\,{\rm d}x, \,\,\,\, L(r) = R \left(1 - \left(\frac{r}{R}\right)^2\right)^{1/2\!},
\label{integral}
\end{equation}
where $x$ is the coordinate along the line of sight and the integration length $L(r)$ decreases toward the boundary ($r \rightarrow
R$), causing the integrated surface density to decline steeply to zero at $r=R$. This geometry-induced radial dependence must be
explicitly accounted for when fitting observed surface density profiles.

Our basic assumption is that the volume density distribution within the filament boundary can be approximated by the Plummer
function
\begin{equation}
\rho(r) = \rho_{\rm C}\left(1+\left(2^{2/\beta\!}-1\right)\left(\frac{2r}{h}\right)^2\right)^{-\beta/2}\,,
\label{volume_density}
\end{equation} 
where $\rho_{\rm C}$ is the axial volume density, $h$ is the full width at half-maximum of the profile, $\beta$ is the density
slope, and the factor $(2^{2/\beta}-1)$ ensures that $h$ remains independent of $\beta$ \citep[][]{Men'shchikov2021a}. The profile
resembles a Gaussian at small radii ($r \lesssim h$) and, for a sufficiently extended filament ($R \gg h$), asymptotically
approaches a power law $\rho(r)\propto r^{-\beta}$ at $r \gg h$ (Fig.~\ref{vsdensprofiles}).

We also assume that the corresponding surface density distribution can be represented by a similar function
\begin{equation}
\Sigma(r) = \Sigma_{\rm C}\left(1+\left(2^{2/\gamma\!}-1\right)\left(\frac{2r}{w}\right)^2\right)^{-\gamma/2} 
\!\left(1-\left(\frac{r}{R}\right)^{\epsilon} \right)^{1/2}\,,
\label{surface_density_fittingfun}
\end{equation} 
where $\Sigma_{\rm C}$ is the filament crest surface density, $w$ and $\gamma$ are the intrinsic half-maximum width and slope,
respectively, and the square-root term with exponent $\epsilon$ describes the dependence of $\Sigma(r)$ on the geometry of the
circular boundary (Eq.~(\ref{integral})). In general, $\epsilon$ depends on both the extent $\xi \equiv R/h$ and slope $\beta$ of
the density distribution in Eq.~(\ref{volume_density}). It follows directly from Eq.~(\ref{integral}) that $\epsilon = 2$ in the
limiting case of uniform density. Examples of the surface density profiles are shown in Fig.~\ref{vsdensprofiles}, together
with the volume density profiles from which they were computed. The direct comparison of the two profile types shows that $\rho(r)$
and $\Sigma(r)$ of a filament of finite extent differ systematically in both slope and width, increasingly so toward shallower
slopes and larger extents.

To ensure full consistency between $\Sigma(r)$ in Eq.~(\ref{surface_density_fittingfun}) and $\rho(r)$ in
Eq.~(\ref{volume_density}), we determined empirical relationships between their parameters through the following procedure.
We
computed surface densities for a grid of numerical models with volume densities from Eq.~(\ref{volume_density}), truncated at $r =
R$ and parameterized by the slope $\beta$ and extent $\xi \equiv R/h$ of the power-law structure. For this purpose, we employed the
code \emph{radmc-3d} by C.\,Dullemond\footnote{\url{http://www.ita.uni-heidelberg.de/~dullemond/software/radmc-3d}} to compute
spherical distributions $\rho(r)$ and integrate them along lines of sight, producing high-resolution surface density images from
which we extracted accurate profiles $\Sigma_{\rm N}(r)$. A single central slice through such a spherical distribution yields
the same $\Sigma(r)$ profile as a slice through the corresponding infinite cylinder with the same $\rho(r)$, so spherical geometry
is computationally simpler while providing the cross-sectional profile we need.

To verify the numerical accuracy of the \emph{radmc-3d} surface density profiles, we performed convergence tests by comparing
images computed with 800, 1600, and 3200 radial zones for representative models spanning the full parameter space ($\beta=$ \{0.5,
2, 9\}, $\xi=$ \{1, 8, 64\}). The maximum relative differences between profiles computed with $N=1600$ and $N=3200$ zones are below
0.5\% for all models, with the largest deviations occurring near the outer boundary for steep profiles ($\beta=9$). The errors
halve when $N$ doubles, confirming reliable first-order convergence. We therefore treat the $N=1600$ profiles as accurate numerical
references for deriving the empirical relationships.

For each numerical model, we then determined the parameters $w$, $\gamma$, and $\epsilon$ of Eq.~(\ref{surface_density_fittingfun})
by adjusting them interactively until the analytical $\Sigma(r)$ visually matched the numerical $\Sigma_{\rm N}(r)$ across the full
radial range. This procedure yielded a discrete set of $(w, \gamma, \epsilon)$ values, one per model. This visual matching is
a one-time calibration step, performed for every grid model solely to construct the analytical relationships
Eqs.~(\ref{betaformula}), (\ref{hHformula}), and (\ref{epsilonformula})--(\ref{coeffshH}). It is distinct from the fitting of
observed profiles (Sect.~\ref{fittingstrategy}), where only $\gamma$, $R$, and $\Sigma_{\rm C}$ are free parameters and $w$,
$\epsilon$, $\beta$, $h$, and $\xi$ are derived from the relationships. The matching reaches high accuracy across nearly the entire
grid; its attainable precision is intrinsically lower only for the extremely shallow and extended models ($\beta\lesssim 0.7$,
$\xi\gtrsim 10$), because $w$, $\gamma$, and $\epsilon$ there become increasingly sensitive functions of $\beta$ and $\xi$
(\chgn{Sect.~\ref{accuracy.analytical}}). We then constructed the analytical relationships Eqs.~(\ref{betaformula}), (\ref{hHformula}), and
(\ref{epsilonformula})--(\ref{coeffshH}) by drawing smooth functions through these discrete values across the full parameter space,
adding terms where needed to improve local accuracy.

The simple form of Eq.~(\ref{surface_density_fittingfun}) and the more elaborate form of Eqs.~(\ref{betaformula}),
(\ref{hHformula}), and (\ref{epsilonformula})--(\ref{coeffshH}) reflect their distinct roles:
Eq.~(\ref{surface_density_fittingfun}) is the surface density model with a small number of physically interpretable parameters,
whereas Eqs.~(\ref{betaformula}), (\ref{hHformula}), and (\ref{epsilonformula})--(\ref{coeffshH}) are empirical interpolation
formulae whose complexity simply reflects the adjustments needed to fit the discrete numerical data to the targeted accuracy.

\begin{figure*}[ht!]
\centerline{\resizebox{0.3249\hsize}{!}{\includegraphics{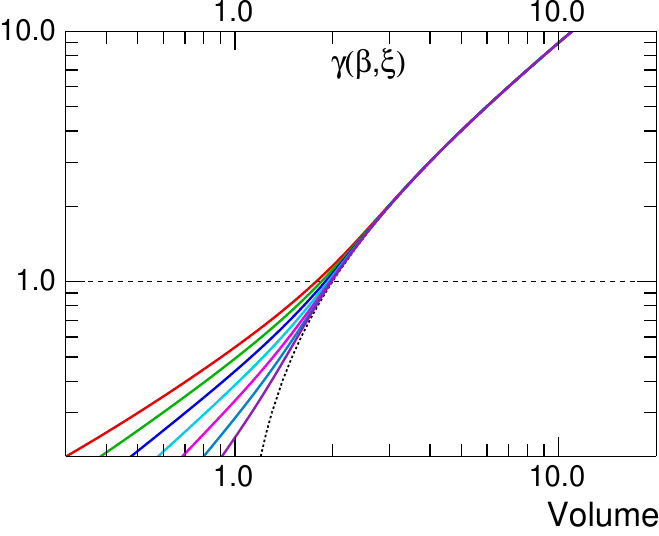}}
            \resizebox{0.3590\hsize}{!}{\includegraphics{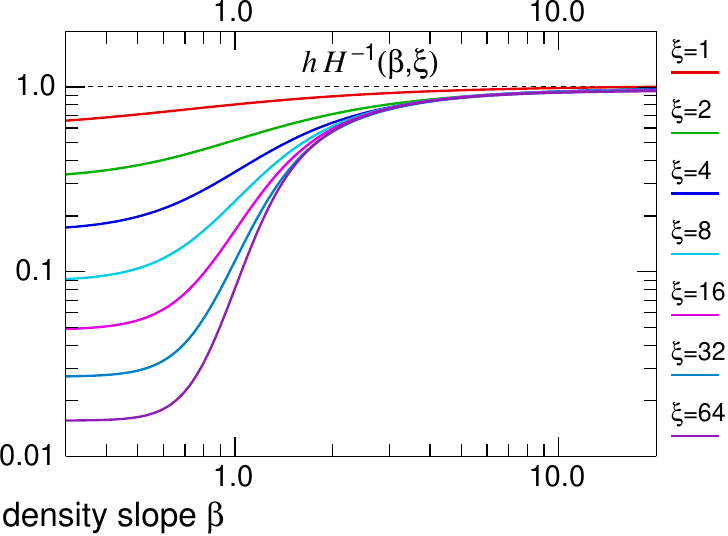}}}
\caption{
\chgn{Analytical dependencies of surface and volume density parameters on $\beta$ for selected values of $\xi$ 
(Eqs.~(\ref{betaformula}) and (\ref{hHformula})): the surface density slope $\gamma$ (\emph{left}) and the ratio of the 
volume and surface density half-maximum widths $hH^{-1}$ (\emph{right}). The dotted curve on the left is the relation 
$\gamma = \beta - 1$ adopted in previous studies, illustrating that this approximation becomes invalid for $\beta \lesssim 2$. 
Dependencies for $\epsilon$ and $w$ are shown in Fig.~\ref{relationships_aux} of Appendix~\ref{app:relationships}.}
} 
\label{relationships}
\end{figure*}

Using the empirical approach described above, we derived analytical relationships between the parameters of the volume and surface
density profiles. The volume density slope $\beta$ can be expressed as a function of the surface density slope $\gamma$ and
the extent $\xi$ as
\begin{align}
\begin{split}
\beta &= \gamma + 
1.529 \left(1 + \xi^{-0.199}\exp\left(-2.725 \left(0.5\gamma\left(\xi^{0.03\gamma^{-0.7}}\right.\right.\right.\right. \\  
&\,\,\,\left.\left.\left.\left. +\,\xi^{0.26\gamma^{-0.1}\xi^{0.03}}\right) -0.319\right)\right)\right)^{-1} - 0.541.
\label{betaformula}
\end{split}
\end{align}
Notably, the difference $\delta\equiv\beta-\gamma$ progressively decreases below unity for $\beta \lesssim 2$ and $\xi \lesssim 10$
(Fig.~\ref{relationships}), in contrast to the fixed value $\delta = 1$ implicitly adopted in previous observational studies that
used Eq.~(\ref{plummerfun}) to derive volume density profiles from surface density observations.

The half-maximum width $h$ of the volume density profile can be approximated as a function of $H$, $\beta$, and $\xi$:
\begin{equation}
h = H \left(S+\left(E-S\right)\left(1+\left(\beta/G\right)^{F}\right)^{-Z}\right),
\label{hHformula}
\end{equation}
where the coefficients $E$, $F$, $G$, $S$, and $Z$ are functions of $\xi$, given by Eq.~(\ref{coeffshH}) in
Appendix~\ref{app:relationships}. For the steepest profiles ($\beta \gtrsim 5$), the half-maximum widths $h$ and $H$ are nearly
identical, whereas for shallower profiles ($\beta \lesssim 5$), the measured width $H$ becomes significantly larger than the
intrinsic volume density width $h$ (Fig.~\ref{relationships}).

Equations~(\ref{betaformula}) and (\ref{hHformula}) provide the two quantities of direct physical interest, $\beta$ and $h$,
which an observer wishes to derive from $\Sigma_{\rm O}(r)$. Neither depends on the boundary exponent $\epsilon$ or the intrinsic
width $w$ of Eq.~(\ref{surface_density_fittingfun}); these two auxiliary quantities enter only the surface density model itself,
which is evaluated at every iteration of the profile fitting. Their analytical approximations (Eqs.~(\ref{epsilonformula}) and
(\ref{wHformula})), together with the coefficients of Eq.~(\ref{hHformula}) (Eq.~(\ref{coeffshH})), are collected in
Appendix~\ref{app:relationships}, which completes the set of relationships needed by the algorithm.

The relationships of Eqs.~(\ref{betaformula}), (\ref{hHformula}), and (\ref{epsilonformula})--(\ref{coeffshH}) enable a
self-consistent determination of all profile parameters. Given the measured values $H$ and $R$ from the observed profile
$\Sigma_{\rm O}(r)$ and treating $\gamma$ as a free fitting parameter, we first substitute $\beta(\gamma,\xi)$ from
Eq.~(\ref{betaformula}) into Eq.~(\ref{hHformula}) and then solve for the extent $\xi \equiv R/h$ numerically using the bisection
method. Once $\xi$ is determined, all remaining parameters ($\beta$, $\epsilon$, $w$, and $h$) are computed directly from
Eqs.~(\ref{betaformula}), (\ref{hHformula}), (\ref{epsilonformula}), and (\ref{wHformula}). This procedure ensures full
consistency between the volume density profile $\rho(r)$ in Eq.~(\ref{volume_density}) and the surface density profile $\Sigma(r)$
in Eq.~(\ref{surface_density_fittingfun}). Although a purely numerical tabulation of the relationships as functions of $\beta$ and
$\xi$ would have been an alternative, any tabulation requires an external interpolation scheme (linear, spline, or similar) to
evaluate the relationships at arbitrary parameter values, with curvature-related errors between the grid points. The analytical
expressions in this paper are, by construction, smooth interpolation functions across the full $(\beta, \xi)$ parameter space, with
maximum residuals from the discrete numerical samples quantified in Table~\ref{accuracytable}. They also provide physical insight
into the limiting behavior of the relationships, and can be implemented compactly in any programming language without requiring
external data files.

\subsection{Fitting strategy}
\label{fittingstrategy}

The fitting procedure was implemented as a \emph{Python} program using the constrained least-squares trust region reflective
(\emph{trf}) algorithm to optimize the parameters of Eq.~(\ref{surface_density_fittingfun}) and minimize the differences between
the fitting function and the observed surface density profile $\Sigma_{\rm O}(r)$. The procedure employs the analytical
relationships from Eqs.~(\ref{betaformula}), (\ref{hHformula}), and (\ref{epsilonformula})--(\ref{coeffshH}) to ensure consistency
between the fitted and derived parameters of the surface and volume density profiles, reducing the number of free fitting
parameters in Eq.~(\ref{surface_density_fittingfun}) to only three: the surface density slope $\gamma$, the boundary radius $R$,
and the crest surface density $\Sigma_{\rm C}$. All other parameters ($\beta$, $\xi$, $\epsilon$, $w$, and $h$) are derived from
these three free parameters using the analytical relationships. We explored several alternative choices of the free
parameters, and the combination $(\gamma, R, \Sigma_{\rm C})$ is the one that keeps the fit best conditioned. In practice, $R$ and
$\Sigma_{\rm C}$ are tightly constrained by their directly measured initial values and are allowed to vary only within narrow
bounds (Sect.~\ref{optimization}). By contrast, the slope $\gamma$ is permitted to vary over a wide range
and is thus effectively the single genuinely free parameter of the fit. In short, the recipe followed by the observer is: apply
Eq.~(\ref{surface_density_fittingfun}) to the observed $\Sigma_{\rm O}(r)$ to obtain $(\gamma, R, \Sigma_{\rm C})$ from the fit,
then derive the physical parameters $(\beta, h)$ through Eqs.~(\ref{betaformula}), (\ref{hHformula}), and
(\ref{epsilonformula})--(\ref{coeffshH}).

The half-maximum width $H$ plays a special role in this scheme. It is not a free fit parameter, but an independent quantity
measured directly from the observed profile, which then enters Eqs.~(\ref{hHformula}) and (\ref{wHformula}) as a fixed input. In
this way $H$ supplies additional information from the data that constrains the model: together with the free slope $\gamma$, it
fixes the extent $\xi$ (through the numerical solution of Eq.~(\ref{hHformula})) and hence all the remaining parameters, which is
what allows the fit to be reduced to only three free parameters. Because $H$ is model-independent and can be read from any observed
or fitted profile, it is also the natural common quantity for comparing methods (Sect.~\ref{comparison_old_method}). The intrinsic
surface density width $w$ is, likewise, not fitted directly but obtained from $H$ through Eq.~(\ref{wHformula}). \chgn{We note that
$H$, as a single measured quantity, is inherently more susceptible to noise and to local profile variations than the parameters
$\gamma$, $R$, and $\Sigma_{\rm C}$, which are constrained by the fit to the entire profile; this is a limitation inherent to any
method that relies on a directly measured width.}

\subsubsection{Initial parameter estimates}
\label{initial_estimates}

The half-maximum width $H$ used in Eqs.~(\ref{wHformula}) and (\ref{hHformula}), as well as the initial estimates $R_{0}$ and
$\Sigma_{\rm C0}$ (subscript~0 denoting values measured directly from the observed profile before fitting), are determined from
$\Sigma_{\rm O}(r)$. The boundary radius $R_{0}$ is estimated independently on each side of the filament at the points where the
profile reaches background levels, and the background of the entire two-sided profile is linearly interpolated and subtracted
before fitting.

Our experience has shown that $\gamma$ is the most sensitive fitting parameter; therefore, we devised a more elaborate scheme for
estimating its initial value $\gamma_{0}$. We create a grid of $\gamma$ values ranging from 0.01 to a sufficiently large
$\gamma_{\rm max\!}$, spaced by a multiplicative factor of 1.05. For each grid value of $\gamma$, we solve Eq.~(\ref{hHformula})
numerically for $\xi$ and calculate $w$ and $\epsilon$ from Eqs.~(\ref{wHformula}) and (\ref{epsilonformula}) using the measured
values $H$ and $R_{0}$. This set of parameters fully defines a trial $\Sigma(r)$ profile via
Eq.~(\ref{surface_density_fittingfun}), which is then compared to the observed $\Sigma_{\rm O}(r)$. The grid value of $\gamma$ that
yields the best match in terms of goodness of fit is chosen as $\gamma_{0}$. Testing of this scheme showed that it often produced
better fitting results than simply using the midpoint between the parameter bounds as $\gamma_{0}$.

\subsubsection{Optimization procedure}
\label{optimization}

The fitting algorithm is constrained with appropriate bounds on the free parameters: $[0.9, 1.1]\,R_{0}$ for $R$, $[0.8,
1.25]\,\Sigma_{\rm C0}$ for $\Sigma_{\rm C}$, and $[0.01, \gamma_{\rm max}]$ for $\gamma$, where we choose $\gamma_{\rm max}$ from
the range [8, 20] depending on the specific application: for non-convolved model profiles we use $\gamma_{\rm max}=20$ to allow the
optimizer to span the full mathematical range of model slopes used in the validation tests; for convolved models and for real
observed filaments we use $\gamma_{\rm max}=9$, because convolution with a Gaussian beam smoothes the profile toward a Gaussian
shape, making steep power-law slopes $\gamma\gtrsim 10$ indistinguishable from a Gaussian shape after convolution, especially in
the presence of noise and background fluctuations typical of real observations. The analogous bound $\beta_{\rm max}$ used in
traditional Plummer fitting (Sect.~\ref{comparison_old_method}) is set to 21 for non-convolved models (Fig.~\ref{modelacc_plummer})
and 10 for convolved models (Fig.~\ref{conv_plummer_accuracies}), following the same reasoning. For real observed filaments we
recommend the range $[0.01, 8]$ for the $\gamma$ bounds. The upper bound $\gamma_{\rm max}=8$ is set empirically: it is the largest
value of $\gamma$ found by the new method when applied to filaments extracted from seven molecular clouds in the companion paper of
\cite{ZhangMenshchikovLi2026}. It is not based on any theoretical argument but serves only to keep the optimizer from drifting to
extreme parameter values in the rare cases where the data are nearly noise-dominated. In each iteration of the optimization
algorithm, the analytical relationships are applied in the following sequence: given the current estimates of $\gamma$ and $R$, (1)
the extent $\xi$ is derived numerically from Eq.~(\ref{hHformula}) using the bisection method, (2) the volume density slope $\beta$
is computed from Eq.~(\ref{betaformula}), (3) the intrinsic surface density width $w$ is determined from Eq.~(\ref{wHformula}), and
(4) the boundary exponent $\epsilon$ is evaluated from Eq.~(\ref{epsilonformula}). These derived parameters, together with the
current value of $\Sigma_{\rm C}$, fully define the model profile $\Sigma(r)$ in Eq.~(\ref{surface_density_fittingfun}), which is
then compared to the observed profile.

For non-convolved model profiles with large dynamic range between the crest and boundary values, we found it beneficial to perform
the fitting in logarithmic space ($\log\Sigma$ vs. $\log r$) to improve both accuracy and robustness. However, our tests showed
that for convolved models, fitting in linear space ($\Sigma$ vs. $r$) recovers more accurate parameters. This suggests that linear
space fitting can also be beneficial when fitting observed profiles affected by convolution with telescope beam.

\subsubsection{Quality assessment}
\label{quality}

Uncertainties of the derived parameters are estimated by the fitting algorithm from the covariance matrix, and the reliability of
each fit is assessed using several diagnostics. These include the goodness of fit (coefficient of determination) $\textsc{R}^2$
$\equiv 1 - S_{\rm res}/S_{\rm tot}$, where $S_{\rm res}$ and $S_{\rm tot}$ are the residual and total sums of squares,
respectively \citep{Draper_Smith1998}, the condition number of the covariance matrix (indicating potential degeneracies), and the
diagonal elements of the covariance matrix corresponding to each fitting parameter (indicating individual parameter uncertainties).
In particular, we require that reliable fits satisfy $\textsc{R}^2 > 0.97$ and have the diagonal element of the covariance matrix
for $\gamma$ below a threshold of $2$, a value found empirically to be appropriate in extensive applications to filaments in seven
molecular clouds \citep{ZhangMenshchikovLi2026}. For comparison, in the same applications the diagonal elements of the covariance
matrix for the other fit parameters $R$ (in pc) and $\Sigma_{\rm C}$ (normalized to unity) are typically of order
$10^{-3}$--$10^{-2}$ and always below $0.1$; the much larger threshold for $\gamma$ reflects the substantially wider range of
conditioning encountered for this parameter. Fits that fail these criteria are discarded. The diagonal element for $\gamma$ can
become large in several circumstances: when the profile is steep (confined to a narrow radial range, so changes in $\gamma$ barely
affect the observed shape), when the profile is compact (small $\xi$, so the boundary term dominates before a power-law range can
develop), or when background fluctuations and noise shorten or eliminate the observable power-law range.


\section{Validation}
\label{validation}

To evaluate the performance and reliability of the fitting method, we conducted extensive numerical tests using model surface
density profiles. Our validation strategy was designed to assess the accuracy of the derived parameters relative to the original
models, the impact of insufficient angular resolution, and the performance when applied to observed profiles.

\subsection{``Infinite'' angular resolution (non-convolved models)}
\label{pure_models}

The most fundamental test of the method presented in Sect.~\ref{method} was conducted to assess its accuracy in the ideal case:
fitting the same numerical model profiles that were used to derive the analytical relationships in Eqs.~(\ref{betaformula}),
(\ref{hHformula}), and (\ref{epsilonformula})--(\ref{coeffshH}). These tests allowed precise evaluation of the accuracy of the
analytical approximations in Eqs.~(\ref{betaformula}), (\ref{hHformula}), and (\ref{epsilonformula})--(\ref{coeffshH}), the
robustness of the fitting strategy, and the accuracy of the derived parameters in comparison with the true input values. The
parameter space of the numerical models sampled the true slopes $\beta_{\rm T} =$ \{0.3, 0.5, 0.6, 0.75, 0.85, 1, 1.25, 1.5, 1.8,
2, 2.5, 3, 4, 6, 9, 18\} and extents $\xi_{\rm T} =$ \{1, 2, 4, 8, 16, 32, 64\} (subscript~T denotes true model values, known only
in the context of numerical tests), providing 112 test cases spanning a wide range of filament morphologies. The range of
$\beta_{\rm T}$ spans from very shallow ($\beta_{\rm T}=0.3$) to very steep ($\beta_{\rm T}=18$, strongly centrally condensed)
profiles, covering all volume density distributions observed or theoretically expected in interstellar filaments. The range of
$\xi_{\rm T}$ spans from compact structures ($\xi_{\rm T}=1$, where the boundary radius equals the half-maximum width) to highly
extended power-law envelopes ($\xi_{\rm T}=64$), encompassing the full range of filament morphologies accessible in \emph{Herschel}
observations.

\begin{figure*}[ht!]
\centerline{\resizebox{0.3300\hsize}{!}{\includegraphics{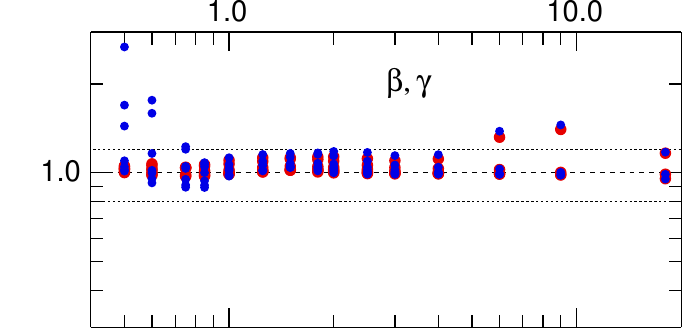}}
            \resizebox{0.3072\hsize}{!}{\includegraphics{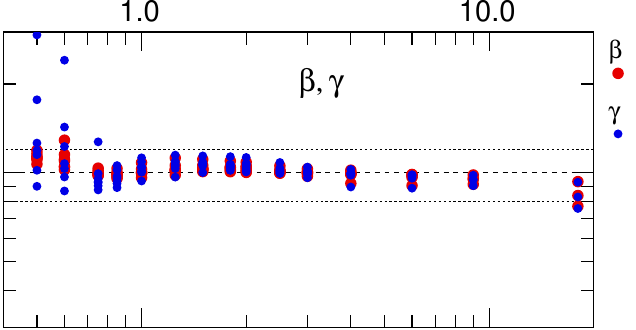}}}
\centerline{\resizebox{0.3300\hsize}{!}{\includegraphics{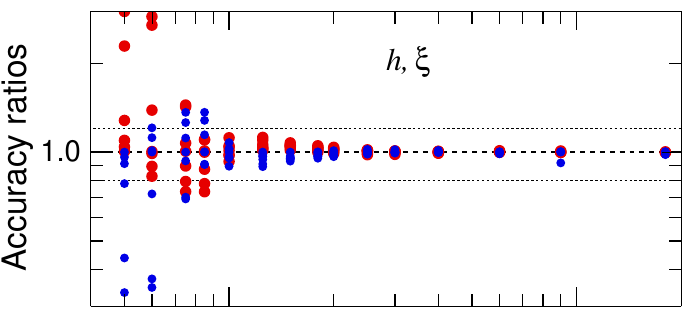}}
            \resizebox{0.3072\hsize}{!}{\includegraphics{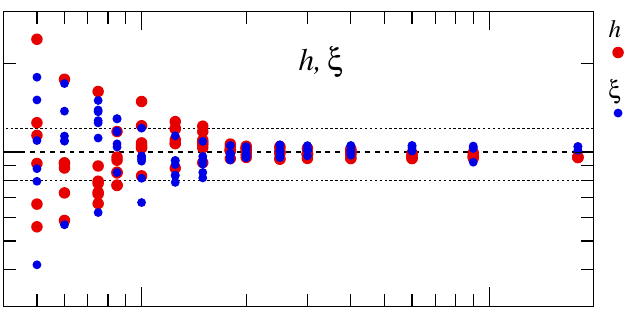}}}
\centerline{\resizebox{0.3300\hsize}{!}{\includegraphics{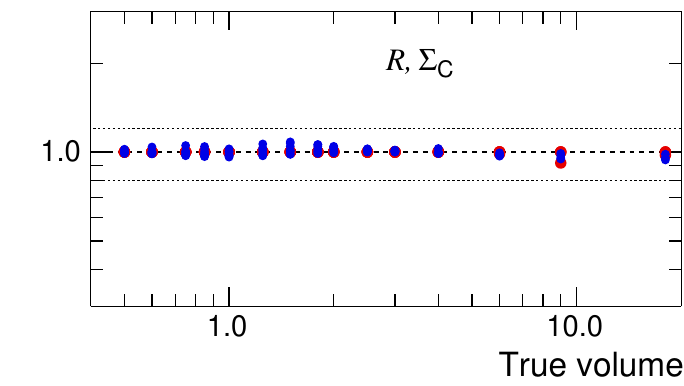}}
            \resizebox{0.3072\hsize}{!}{\includegraphics{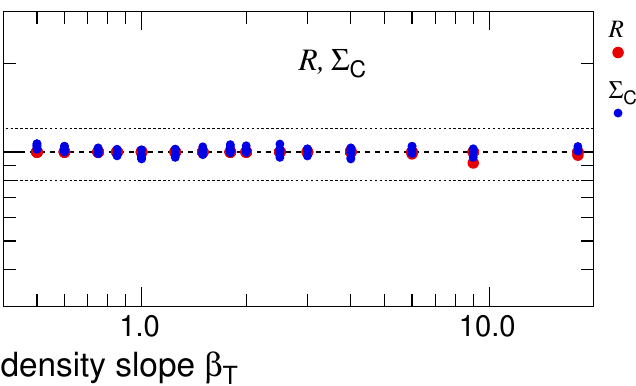}}}
\caption{
\chgn{Accuracies of the main parameters derived from fitting the numerical model profiles.} The accuracy ratios with respect to the 
true model values are shown as functions of the true $\beta_{\rm T}$ for fitted surface density profiles without noise (\emph{left}) 
and with 10\% noise added (\emph{right}). Filled circles represent the parameters indicated at the edge of the right panels; their
vertical spread reflects variations due to different $\xi_{\rm T}$ values. Horizontal dashed lines indicate perfect accuracy, while
dotted lines mark $\pm 20\%$ error bounds. The fitting (in logarithmic space) was constrained with an upper bound of 
$\gamma_{\rm max\!}=20$. \chgn{Accuracies of the parameters $\epsilon$ and $w$ are shown in Fig.~\ref{modelacc_aux} of 
Appendix~\ref{app:accuracy}.}
} 
\label{modelacc}
\end{figure*}

\subsubsection{Accuracy of analytical approximations}
\label{accuracy.analytical}

To assess the intrinsic accuracy of the analytical relationships, we first compared the analytical surface density profiles from
Eq.~(\ref{surface_density_fittingfun}) (using the true model parameters) to the true numerical profiles $\Sigma_{\rm N}(r)$ in
terms of the maximum and median relative deviations over all radial points. We also evaluated the relative errors in the parameters
computed from the analytical relationships Eqs.~(\ref{betaformula}), (\ref{hHformula}), and
(\ref{epsilonformula})--(\ref{coeffshH}) with respect to the true model parameters.

Both the analytical profiles and the derived parameters reproduce the numerical models accurately across nearly the entire
parameter space, with typical errors below a few percent. Larger deviations occur only for very shallow slopes
($\beta_{\rm T}\lesssim 0.7$) and extended profiles ($\xi_{\rm T}\gtrsim 10$), where the visual matching is intrinsically less
precise (Sect.~\ref{fittingfun}); such extreme parameter combinations are unlikely to occur in practice. The detailed residual
curves, for the analytical profiles and for each of the derived parameters, are presented in Appendix~\ref{app:relationships}
(Fig.~\ref{relerrors}), alongside the relationships whose accuracy they demonstrate. The primary statement of accuracy,
for the two parameters of direct physical interest, is given in Table~\ref{accuracytable}.

Table~\ref{accuracytable} summarizes the maximum relative errors in the parameters $\beta$ and $h$ derived from the analytical
relationships, as functions of the fitted parameter ranges, for noise-free profiles. Across the full tested range
$\beta_{\rm T}\in[0.75, 9]$ and $\xi_{\rm T}\in[1, 64]$, the maximum errors are $1.6\%$ in $\beta$ and $7\%$ in $h$; they drop
rapidly as the lower bound on $\beta_{\rm T}$ is raised. For the median observed slopes $\beta\approx 2.1$--$2.4$ found by
\cite{ZhangMenshchikovLi2026} in seven molecular clouds, the formulae recover $\beta$ to within $0.01\%$ and $h$ to within
$2\%$.

\begin{table}
\caption{Maximum relative errors in the parameters $\beta$ and $h$ derived from the analytical relationships
(Eqs.~(\ref{betaformula}), (\ref{hHformula}), and (\ref{epsilonformula})--(\ref{coeffshH})), as functions of the fitted 
parameter ranges, for noise-free profiles.}
\label{accuracytable}
\centering
\begin{tabular}{cccc}
\hline\hline
\noalign{\smallskip}
$\beta_{\rm T}$ range & $\xi_{\rm T}$ range & $\Delta\beta/\beta$\,\,\,\, & $\Delta h/h$\\
\noalign{\smallskip}
\hline
\noalign{\smallskip}
$0.75$--$9.0$ & $1.0$--$64.0$ & $0.0158\,\,\,\,\,\,$ & $0.0700$\\
$1.00$--$9.0$ & $1.0$--$64.0$ & $0.00278\,\,\,$      & $0.0197$\\
$1.50$--$9.0$ & $1.0$--$64.0$ & $0.000700$           & $0.0197$\\
$2.00$--$9.0$ & $1.0$--$64.0$ & $0.000104$           & $0.0194$\\
\noalign{\smallskip}
\hline
\end{tabular}
\end{table}

\subsubsection{Fitting accuracy with and without noise}

The overall fidelity of the method was verified by fitting the model profiles and comparing the derived parameters to the true
input values (Fig.~\ref{modelacc}). Note that the errors in Fig.~\ref{modelacc} may exceed those in Fig.~\ref{relerrors}, because
the two figures measure different things. Fig.~\ref{relerrors} evaluates the analytical relationships Eqs.~(\ref{betaformula}),
(\ref{hHformula}), and (\ref{epsilonformula})--(\ref{coeffshH}) alone: they are evaluated at the true input $(\beta_{\rm T},
\xi_{\rm T})$ and compared with the per-model values obtained by visual matching (Sect.~\ref{fittingfun}); the only error source
is the irreducible deviation of these formulae from the discrete per-model fits. Fig.~\ref{modelacc}, in contrast, tests the
complete fitting pipeline of Sect.~\ref{fittingstrategy}, in which $\gamma$, $R$, and $\Sigma_{\rm C}$ are fitted as free
parameters of Eq.~(\ref{surface_density_fittingfun}), and $\beta$ and $h$ are subsequently derived through
Eqs.~(\ref{betaformula}), (\ref{hHformula}), and (\ref{epsilonformula})--(\ref{coeffshH}). Its errors therefore include both the
formula residuals and the additional uncertainty from parameter sensitivity, degeneracies, and the finite precision of the
numerical optimization. In addition to the original noise-free profiles, we also tested more realistic ones whose points were
modified by random Gaussian noise at levels of 5, 10, 20, and 30\% relative to the crest surface density. Without noise, the
derived parameters achieve accuracies within a few percent for $\beta_{\rm T} > 0.85$, whereas for shallower density slopes, the
fitting accuracies are somewhat degraded due to the limitations of the analytical approximations discussed above
(Fig.~\ref{modelacc}). As expected, fitting the model profiles with added 10\% random noise reveals some degradation in the derived
parameters, primarily for $\beta_{\rm T}\lesssim 1.5$, where the shallower profiles are more sensitive to noise contamination. At
higher noise levels (20--30\%), the fitting becomes increasingly challenging for all but the steepest profiles ($\beta_{\rm T}
\gtrsim 3$), though the method remains robust in most cases.

The two error contributions dominate in different regimes. The residuals quantified in Table~\ref{accuracytable} set the error
floor and dominate for shallow profiles ($\beta_{\rm T}\lesssim 1$), where they reach a few percent. For steep profiles
($\beta_{\rm T}\gtrsim 6$), by contrast, the formula residuals are negligible (Table~\ref{accuracytable}) and the larger scatter
visible in Fig.~\ref{modelacc} (e.g., at $\beta_{\rm T}=6$ and $9$) is dominated by the fitting uncertainty: steep profiles occupy
a narrow radial range, so $\gamma$ is poorly conditioned and the optimization is more sensitive to the discretized profile sampling
(Sect.~\ref{quality}). At intermediate slopes both contributions are small.

The test results presented in Figs.~\ref{relerrors} and \ref{modelacc} demonstrate that the method achieves high accuracy across a
wide range of filament parameters under ideal conditions (infinite angular resolution), effectively defining its applicability
domain and establishing a baseline for more realistic tests with finite angular resolution.

\subsection{Variable angular resolutions (convolved models)}
\label{convolved_models}

The performance of the method was also tested under more realistic and challenging conditions similar to those encountered in
observational studies. Fitting observed surface density profiles relies on the assumption that the telescope beam width $O$ is much
smaller than the characteristic widths $H$ of the observed structures. Effects of insufficient angular resolution on measured
properties can be very strong \citep{Men'shchikov2023}; therefore, they must be carefully examined using convolved model profiles.

\subsubsection{Model setup and resolvedness range}

We created models of a long straight filament using the $\rho(r)$ profiles from Eq.~(\ref{volume_density}) with a half-maximum
width $h_{\rm T}=147.33$\arcsec, slopes $\beta_{\rm T} =$ \{0.5, 1, 2, 3, 6, 9\}, and extents $\xi_{\rm T}\equiv R_{\rm T}/h_{\rm
T} =$ \{1, 1.41, 2, 2.83, 4, 5.66, 8\}. The corresponding surface density images were computed from
Eq.~(\ref{surface_density_fittingfun}) with $\gamma$, $w$, and $\epsilon$ evaluated from
Eqs.~(\ref{betaformula}), (\ref{wHformula}), and (\ref{epsilonformula}) using the known true values
$\beta_{\rm T}$ and $\xi_{\rm T}$, ensuring full consistency with the volume density models, so
that any deviations in subsequent fitting tests reflect only the effects of finite angular resolution.
These images were convolved with Gaussian kernels of half-maximum widths $O =$ \{3.375, 6.75, 13.5, 27,
54, 108, 216, 432, 864, 1728\}\arcsec. This enabled us to explore a wide range of resolvedness $\mathcal{R}\equiv H/O$
\citep{Men'shchikov2023}, where $H$ is the measured half-maximum width of the convolved surface density profiles. The resulting
values $1< \mathcal{R}\lesssim 100$ covered all cases of interest, from completely unresolved ($\mathcal{R} \approx 1$) to fully
resolved ($\mathcal{R} \gg 10$) filaments. By construction $\mathcal{R}\geq 1$,
since the observed surface density width $H$ cannot be smaller than the beam width~$O$.

The convolved model profiles (especially the steeper ones) extend down to values many orders of magnitude below their peaks.
Convolution spreads the model profile beyond its true boundary radius $R$ and transforms it into a more Gaussian-like shape, with
the most significant changes occurring at very low surface density values many orders of magnitude below $\Sigma_{\rm C}$. However,
fluctuating backgrounds in molecular clouds where filaments form make such low profile values unmeasurable in real observations. To
test the fitting method under such realistic conditions, we also applied it to modified model profiles that were truncated at the
$0.1\Sigma_{\rm C}$ level and contaminated by 10\% random Gaussian noise.

This numerical convolution test serves as the direct assessment of beam impact on the recovery of the physical parameters: the
model profiles produced by Eq.~(\ref{surface_density_fittingfun}) are convolved with Gaussian beams of varying width and refit with
the full pipeline, so that the resulting deviations of $\beta$ and $h$ from their true values quantify the beam-induced bias. The
test inherits the residual errors of the analytical approximations Eqs.~(\ref{betaformula}), (\ref{hHformula}), and
(\ref{epsilonformula})--(\ref{coeffshH}) shown in Figs.~\ref{relerrors} and~\ref{modelacc} (and quantified in
Table~\ref{accuracytable}); the recovered physical parameters carry these residuals as an additional contribution beyond the
convolution-induced error.

\subsubsection{Fitting accuracy as a function of resolvedness}

\begin{figure*}[ht!]
\centerline{\resizebox{0.3090\hsize}{!}{\includegraphics{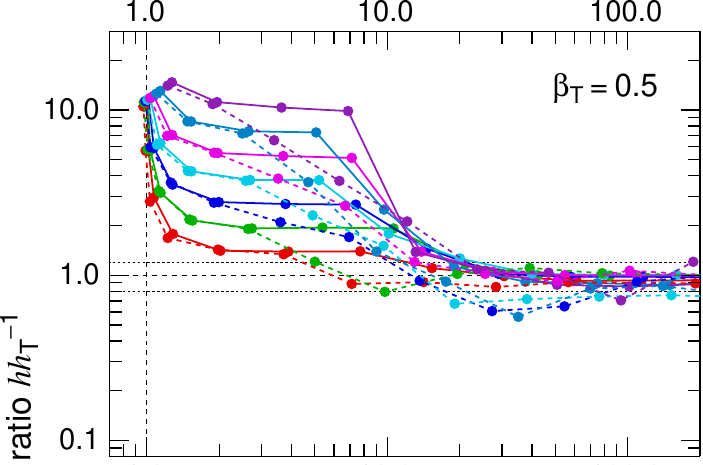}}
            \resizebox{0.2614\hsize}{!}{\includegraphics{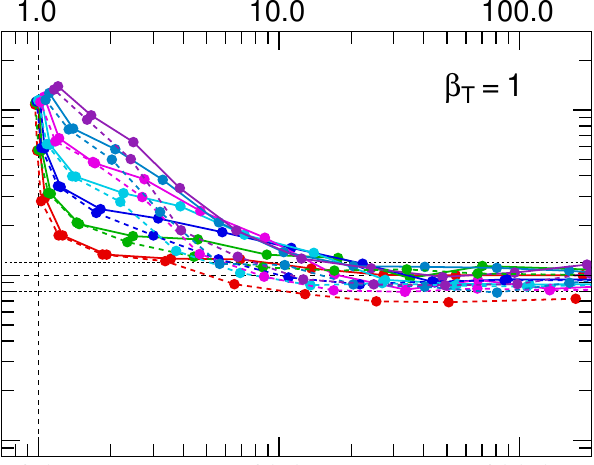}}
            \resizebox{0.3053\hsize}{!}{\includegraphics{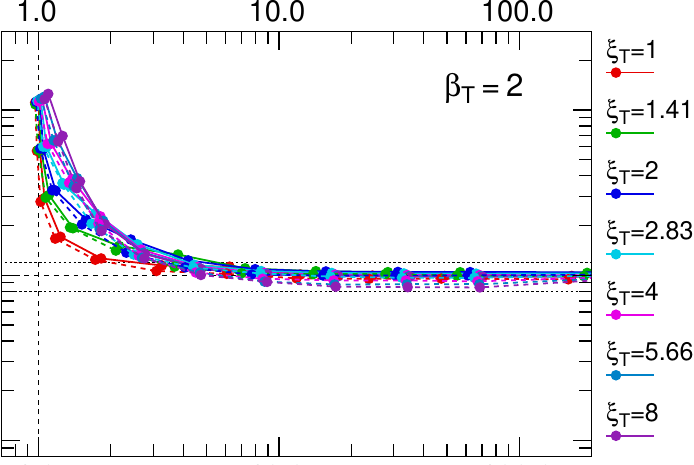}}}
\centerline{\resizebox{0.3090\hsize}{!}{\includegraphics{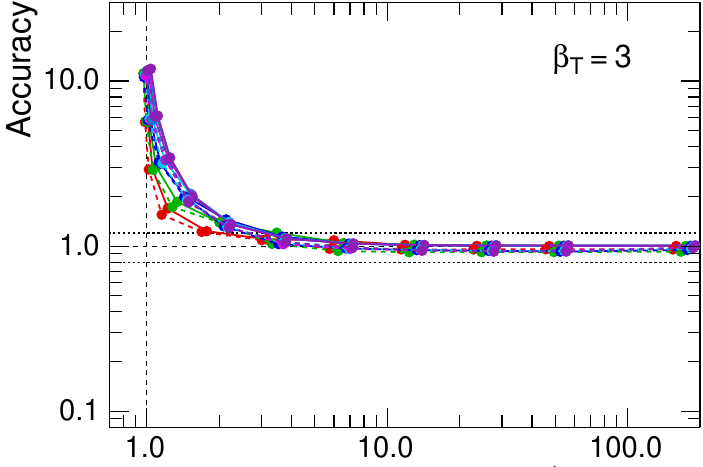}}
            \resizebox{0.2614\hsize}{!}{\includegraphics{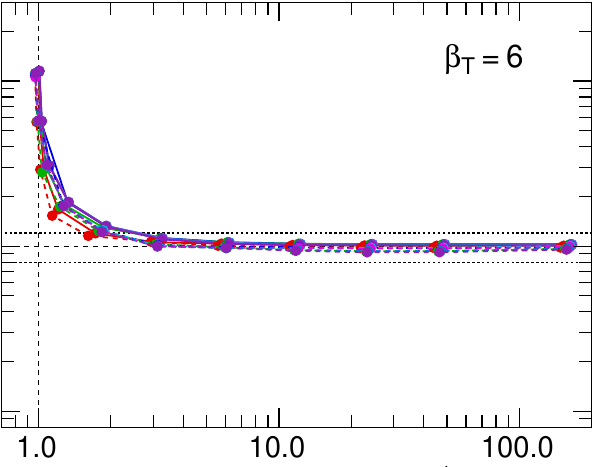}}
            \resizebox{0.3053\hsize}{!}{\includegraphics{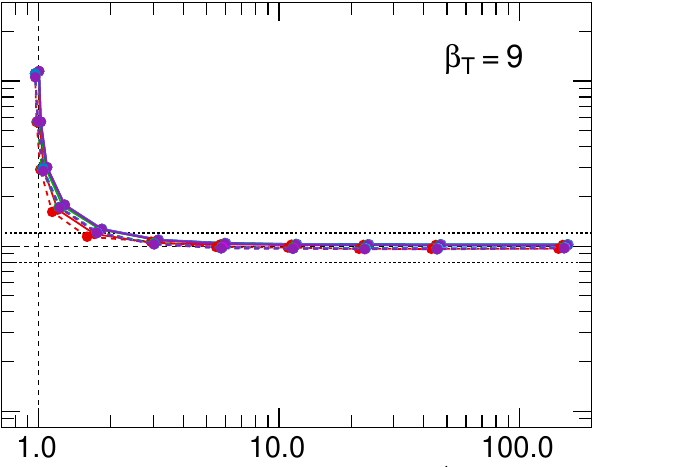}}}
\centerline{\resizebox{0.3090\hsize}{!}{\includegraphics{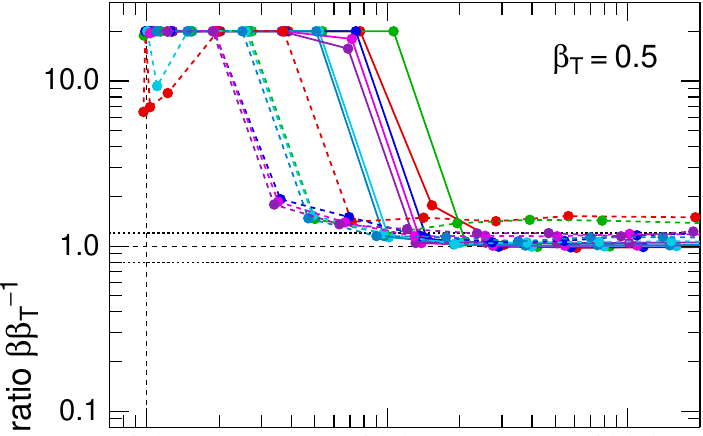}}
            \resizebox{0.2614\hsize}{!}{\includegraphics{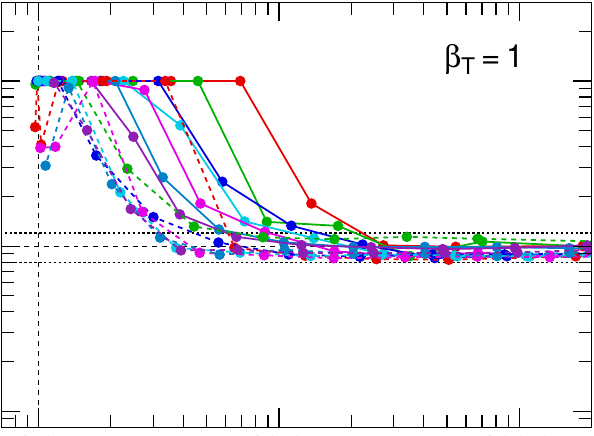}}
            \resizebox{0.3053\hsize}{!}{\includegraphics{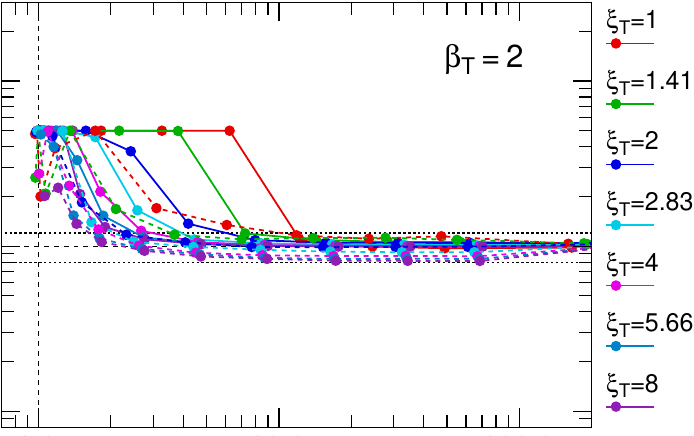}}}
\centerline{\resizebox{0.3090\hsize}{!}{\includegraphics{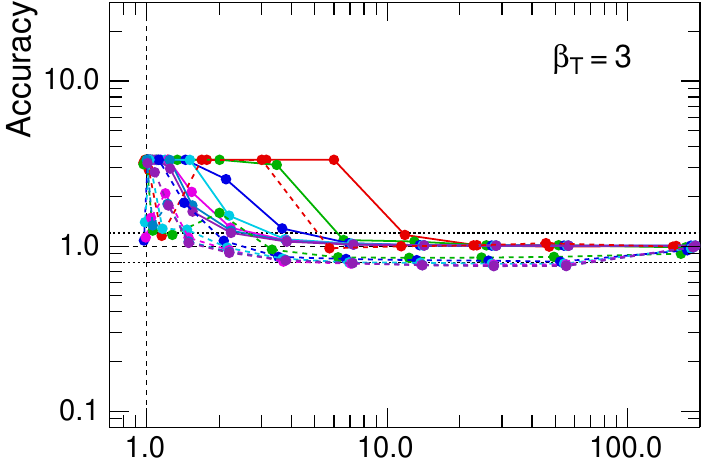}}
            \resizebox{0.2614\hsize}{!}{\includegraphics{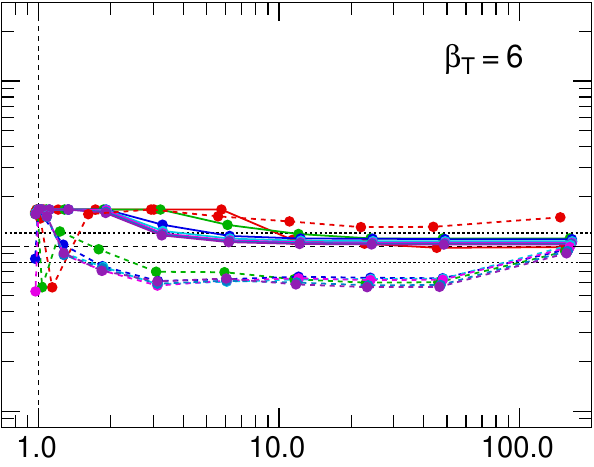}}
            \resizebox{0.3053\hsize}{!}{\includegraphics{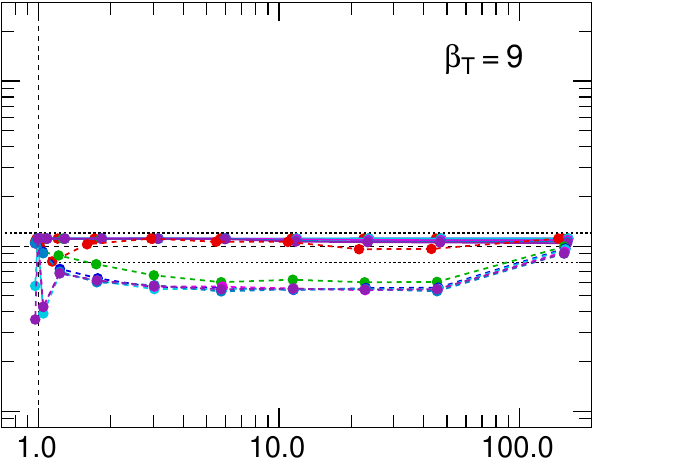}}}
\centerline{\resizebox{0.3090\hsize}{!}{\includegraphics{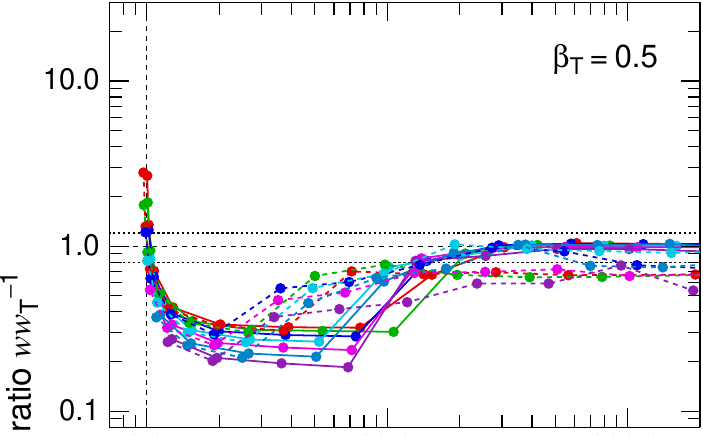}}
            \resizebox{0.2614\hsize}{!}{\includegraphics{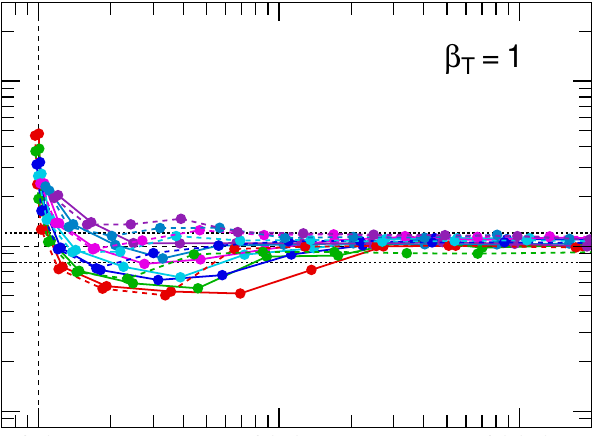}}
            \resizebox{0.3053\hsize}{!}{\includegraphics{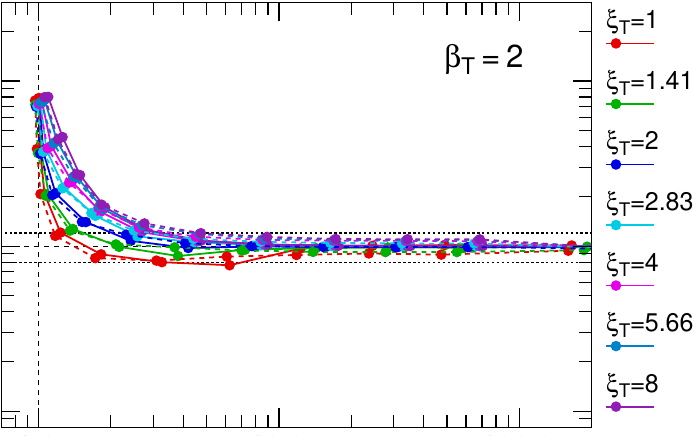}}}
\centerline{\resizebox{0.3090\hsize}{!}{\includegraphics{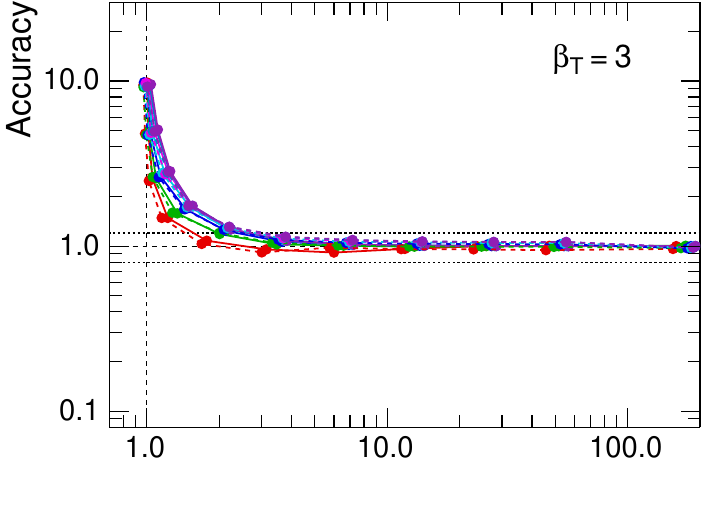}}
            \resizebox{0.2614\hsize}{!}{\includegraphics{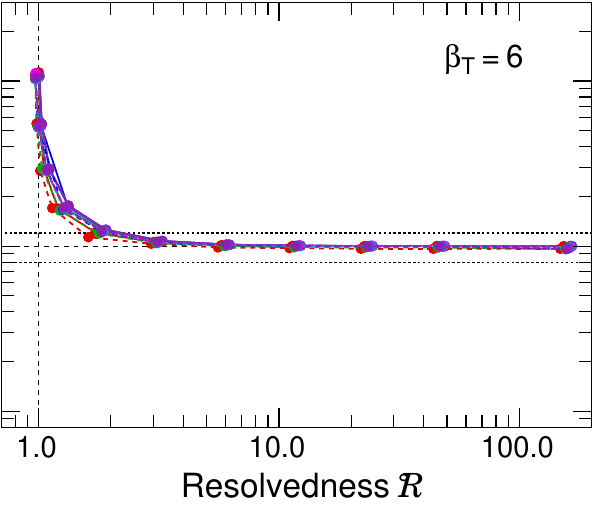}}
            \resizebox{0.3053\hsize}{!}{\includegraphics{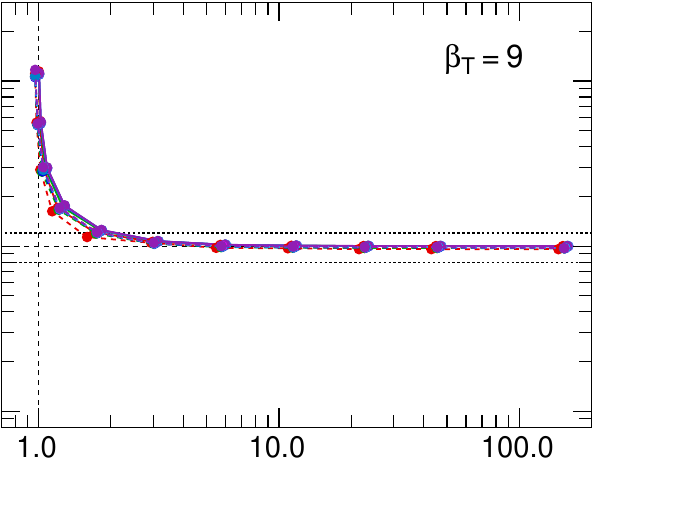}}}
\caption{
Accuracy of the widths and slopes derived from fitting surface density profiles of a model filament convolved with increasingly
larger beams. The accuracy ratios $hh_{\rm T}^{-1}$, $\beta\beta_{\rm T}^{-1}$, and $ww_{\rm T}^{-1}$ are represented by filled
circles connected by solid lines as functions of the profile resolvedness $\mathcal{R}$ for different values of $\beta_{\rm T}$ and
$\xi_{\rm T}$. Dashed lines show accuracies for modified profiles truncated at the $0.1\Sigma_{\rm C}$ level and altered by 10\%
random noise. Vertical dashed lines mark the lower bound for $\mathcal{R}$ values, horizontal dashed lines indicate perfect
accuracy, and dotted lines show $\pm 20\%$ error bounds. The slope accuracy for $\beta_{\rm T\!}=9$ is nearly perfect because the 
fitting (in linear space) was constrained with an upper bound of $\gamma_{\rm max\!}=9$.
} 
\label{conv.accuracies}
\end{figure*}

Fitting results for the convolved model profiles (Fig.~\ref{conv.accuracies}) quantify the resolvedness
assumption stated in Sect.~\ref{convolved_models}: reliable parameter recovery requires the filament to be well resolved. The
precise resolvedness threshold depends on the parameter ($h$, $\beta$, or
$w$) and on the true slope $\beta_{\rm T}$, but as a practical rule all parameters are recovered to within $\sim 20\%$ only for
$\mathcal{R}\gtrsim 20$ (relaxing to $\mathcal{R}\gtrsim 8$ for the truncated, noisy profiles; Sect.~\ref{truncation_noise}).
Models convolved to lower angular resolutions reveal strong systematic biases in the
derived parameters at lower resolvedness, most pronounced for shallower $\rho(r)$ distributions with $\beta_{\rm{T}}\lesssim 2$.
For such slopes, both $\beta$ and $h$ become increasingly overestimated, though with different dependencies on $\xi_{\rm T}$.

For the shallowest slope ($\beta_{\rm{T}}=0.5$), the filament width $h$ remains fairly accurate down to $\mathcal{R}\approx 2$ only
for the least extended filaments ($\xi_{\rm T}\approx 1$), being increasingly overestimated in proportion to $\xi_{\rm T}$
(Fig.~\ref{conv.accuracies}). In contrast, the derived slopes $\beta$ remain relatively accurate for the most extended filaments
with steeper $\rho(r)$ profiles, while their errors for the most compact filaments with shallow slopes ($\beta_{\rm{T}}\lesssim 2$)
increase dramatically at low resolvedness.
Even at high resolvedness ($\mathcal{R}\gtrsim 20$), however, the recovery of $\beta$ for the steepest profiles is not
guaranteed: a steep profile occupies a narrow radial range, so $\gamma$ (and hence $\beta$) is intrinsically poorly conditioned
(Sect.~\ref{quality}), and the recovered value depends sensitively on the upper bound $\gamma_{\rm max}$ imposed in the fit. The
apparently accurate slope for $\beta_{\rm T}=9$ in Fig.~\ref{conv.accuracies} is in fact a consequence of the bound
$\gamma_{\rm max}=9$ used there, rather than an independent confirmation of accurate recovery.

The intrinsic surface density width $w$ exhibits more complex behavior. It is significantly underestimated (by factors of 3--5) at
intermediate resolvedness ($\mathcal{R} \lesssim 20$) for shallow slopes (Fig.~\ref{conv.accuracies}), primarily due to the strong
overestimation of $\beta$ at these resolutions. However, as filaments become fully unresolved ($\mathcal{R} \to 1$), $w$ becomes
progressively overestimated, like all other width parameters. For steeper slopes ($\beta_{\rm T} \gtrsim 2$), $h$ and $w$ behave
nearly identically. We do not show the accuracy of the intrinsic surface density slope $\gamma$ in Fig.~\ref{conv.accuracies}, as
it exhibits essentially the same behavior as $\beta$, though with slightly larger maximum errors (up to a factor of two).

\subsubsection{Effects of truncation and noise}
\label{truncation_noise}

The most important practical regime is $\mathcal{R}\approx 2{-}10$, corresponding to typical
\emph{Herschel} resolvedness values for nearby star-forming regions.
The modified profiles with added 10\% noise and truncated at a relatively high level of $0.1\Sigma_{\rm C}$ show somewhat different
behavior (Fig.~\ref{conv.accuracies}). On one hand, truncation reduces the extent of the convolved profile beyond the true boundary
and the resulting overestimation of the derived radius $R$. Furthermore, it significantly reduces the overestimation of $h$ and
$\beta$ in the practically important range of $5\lesssim \mathcal{R}\lesssim 20$ for shallow slopes ($\beta_{\rm T}\lesssim 2$). On
the other hand, the truncation leads to underestimated $\beta$ and $R$ for steep slopes ($\beta_{\rm T}\gtrsim 2$) and to
underestimated width $h$ for the shallowest slope ($\beta_{\rm T}=0.5$) at high resolvedness ($\mathcal{R}\gtrsim 10$). These
results indicate that profile truncation at observable levels can partially mitigate some systematic biases at low resolvedness but
may introduce new biases for steep profiles at high $\mathcal{R}$.

\subsection{Application to observed filaments}
\label{applied_to_California}

To demonstrate a practical application of the fitting method, we selected surface density profiles of three filaments in the
\object{California} star-forming region observed with \emph{Herschel} \citep{Harvey_etal2013, Zhang_etal2024}. The surface density
images with an angular resolution of $O=13.5$\arcsec\ were derived using the \emph{hires} algorithm, and the filamentary
structures were extracted with the \emph{getsf} method\footnote{\url{http://irfu.cea.fr/Pisp/alexander.menshchikov/}}
\citep{Men'shchikov2021a}. A detailed description of the observational data and filament extraction is presented by
\cite{ZhangMenshchikovLi2026}.

\begin{figure}
\centering
\centerline{\resizebox{0.9\hsize}{!}{\includegraphics{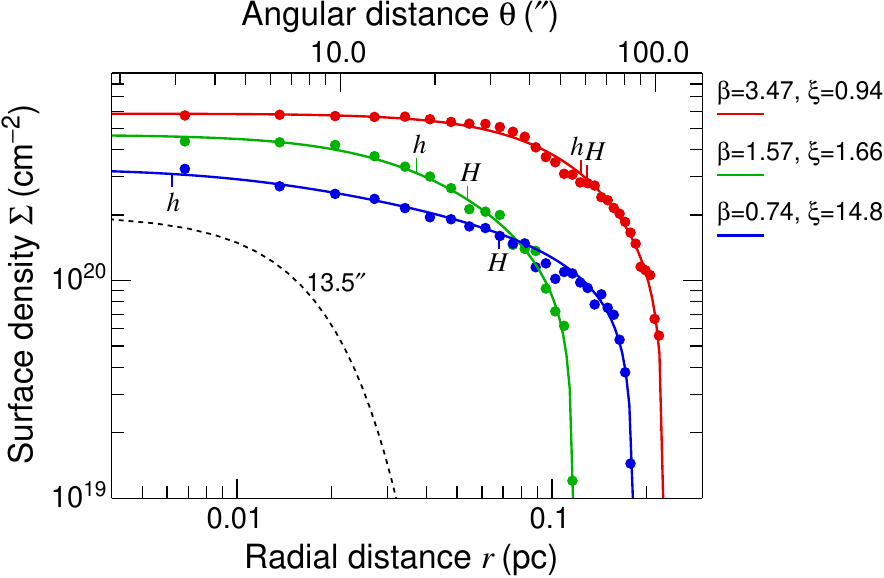}}}
\caption{
Application of the fitting method to three filaments in the \object{California} molecular cloud. The surface density profiles
$\Sigma_{\rm O}(r)$ of selected filaments and the corresponding fitted profiles are shown with dots and solid curves, respectively.
The derived $\beta$ and $\xi$ of the fits are specified, and the half-maximum widths $H$ and $h$ of the $\Sigma(r)$ and $\rho(r)$
profiles, respectively, are indicated next to the curves at the radial positions of their half-widths. The dashed curve displays
the beam profile with half-maximum width $O=13.5$\arcsec.
}
\label{fig:california_profiles}
\end{figure}

\subsubsection{Fitting results}
\label{fitting_results}

The fitting results for the selected California filaments are shown in Fig.~\ref{fig:california_profiles}. All fits were performed
in linear space ($\Sigma$ vs. $r$), as appropriate for beam-convolved profiles (Sect.~\ref{fittingstrategy}). No beam convolution
of the model profile is applied; the analytical $\Sigma(r)$ is fitted directly to the observed profile, under the assumption that
the profiles are sufficiently resolved for beam effects on the profile shape to be negligible. The three surface density profiles
with measured half-maximum widths $H=$ \{0.258, 0.109, 0.137\} pc display very good quality fits (with $\textsc{R}^2=$ \{0.987,
0.993, 0.991\} for the steep, intermediate, and shallow profiles, respectively), yielding derived surface density parameters $w=$
\{0.289, 0.150, 0.397\} pc and $\gamma=$ \{2.49, 0.830, 0.228\}, and corresponding volume density parameters $h=$ \{0.248, 0.0743,
0.0125\} pc, $\beta=$ \{3.47, 1.57, 0.740\}, and $\xi=$ \{0.938, 1.66, 14.8\}. The formal uncertainties on the fitted slopes are
$\sigma_\gamma=$ \{1.20, 0.107, 0.029\} and on the boundary radii $\sigma_R=$ \{0.000, 0.001, 0.002\} pc for the steep,
intermediate, and shallow profiles, respectively. The large $\sigma_\gamma$ for the steep profile arises for the reasons discussed
in Sect.~\ref{quality}. Propagating these fit uncertainties through the analytical chain Eqs.~(\ref{betaformula}),
(\ref{hHformula}), and (\ref{epsilonformula})--(\ref{coeffshH}) (with $\Sigma_{\rm C}$ not entering, since $\beta$ and $h$
depend only on the fitted $\gamma$ and $R$ ($\xi\equiv R/h$) and on the measured $H$, not on the crest normalization) yields the
formal uncertainties on the derived physical parameters: $\sigma_\beta=$ \{1.22, 0.165, 0.021\} and $\sigma_h=$ \{0.0065, 0.0041,
0.0040\} pc for the same three profiles, respectively. Besides, the resolvedness values $\mathcal{R}=$ \{8.38, 3.54, 4.45\}
fall in the range where the model tests of Fig.~\ref{conv.accuracies} indicate large systematic biases, particularly for the second
and third filaments with lower $\mathcal{R}$ and shallower slopes. The associated systematic errors in $\beta$ and $h$ are
therefore expected to exceed the formal $\sigma_\beta$ and $\sigma_h$ above, plausibly by a factor of a few, even though an
accurate value cannot be given without independent measurements of the true filament properties. The formal uncertainties should
thus be regarded as lower bounds on the true parameter errors.

Beyond these convolution-induced systematics, the dominant source of systematic uncertainty in fitting real observed profiles
is background subtraction. Its magnitude is not amenable to a single general estimate: backgrounds vary irreducibly in intensity,
spatial structure, and effective angular scale from cloud to cloud and from filament to filament, and depend additionally on the
blending of nearby structures and on the angular resolution. Linear interpolation across the filament cross-section is the
prevailing approach used in the field, including in the companion paper of \cite{ZhangMenshchikovLi2026}; it is rough but no
general improvement is currently available.

\subsubsection{Implications for observational studies}

Figure~\ref{fig:california_profiles} illustrates two qualitative trends that are frequently overlooked in observational studies of
filaments. First, the half-maximum widths $H$ and $h$ of the surface and volume density profiles, respectively, become very
dissimilar for extended filaments with large $\xi$ and shallow slopes $\beta$. For the three California filaments, the ratios are
$Hh^{-1}=$ \{1.04, 1.47, 11.0\}, with the third filament showing a dramatic difference. Second, the slope differences
$\delta\equiv\beta-\gamma$ are systematically below unity ($\delta=$ \{0.98, 0.74, 0.51\}) for these filaments.
We stress that these particular numbers are illustrative rather than definitive: the three filaments have $\mathcal{R}=$
\{8.38, 3.54, 4.45\}, so the derived $\beta$ and $h$ (and hence the precise $Hh^{-1}$ and $\delta$ values) carry the large systematic
uncertainties discussed in Sect.~\ref{fitting_results}. The qualitative conclusions -- that $H$ and $h$ can differ greatly
and that $\delta$ can fall well below unity for shallow, extended filaments -- are robust and follow directly from the analytical
relationships (Fig.~\ref{relationships}), but the specific values for these particular underresolved filaments should not be taken
as accurate measurements.

These results demonstrate that it is incorrect to measure $H$ and $\gamma$ from a surface density profile and assume that the
volume density distribution has the same (or similar) width $h\approx H$ and slope $\beta=\gamma+1$. The latter relation,
traditionally used to convert a measured surface density slope $\gamma$ to the volume density slope, holds only for sufficiently
steep profiles with $\gamma\gtrsim 1.3$ and moderate extents. For shallower slopes, especially for extended profiles with
$\xi\gtrsim 10$, the traditional assumptions become increasingly inaccurate, as shown in Fig.~\ref{relationships}. The third
California filament, with its shallow slope ($\beta=0.740$) and large extent ($\xi=14.8$), exemplifies this issue, with the
measured width $H$ being more than an order of magnitude larger than the derived volume density width $h$.


\section{Discussion}
\label{discussion}

The new method enables accurate derivation of filament surface and volume density parameters assuming a nonuniform cylindrical
geometry (e.g., a curved cylinder with varying circular cross-sections) and favorable observational conditions -- including the
filament axis being orthogonal to the line of sight, high contrast above background, low background fluctuations, and high angular
resolution. However, in practice, filament geometries and orientations are largely unknown, and observational conditions typically
fall far short of these ideals.

\subsection{Effects of angular resolution on derived parameters}
\label{angular_resolution}

The ranges of relatively accurate parameter derivation revealed by our tests exhibit several important patterns
(Fig.~\ref{conv.accuracies}).

\subsubsection{Volume density width}

Accurate recovery of the volume density width $h$ requires very high resolvedness ($\mathcal{R}\gtrsim 10$) for filaments with
shallow slopes ($\beta_{\rm T}\approx 0.5$), especially for extended ones ($\xi_{\rm T}>1$). For steeper slopes, accurate recovery
of $h$ becomes possible at much lower resolvedness, down to $\mathcal{R}\approx 2{-}3$. This behavior arises because convolution of
shallow filament profiles extends the convolved peaks over much wider areas than convolution of steeper profiles. This is the same
effect that shifts the lower bound of resolvedness $\mathcal{R}_{\rm min}$ from unity to much larger values in
Fig.~\ref{deconvolinf} for ``infinite'' filaments. Contrary to the traditional assumption that structures are resolved when $H
\gtrsim 2O$, which addresses only whether the observed surface density width is resolved, not whether the underlying volume density
structure (characterized by $h\ll H$ for shallow profiles) is physically resolved, the shallowest filaments remain physically
unresolved even when $H\approx 10\,O$ (see Appendix~\ref{deconvolution} for details). We note that the resolvedness $\mathcal{R} =
H/O$ characterizes, by definition, only how the half-maximum width $H$ of the surface density profile compares with the beam $O$;
it does not involve the volume density quantities $h$ or $\xi$. It is therefore distinct from the question of whether the
underlying volume density structure is physically resolved (discussed above), and is also separate from the recoverability of the
filament boundary radius $R$. The boundary of $\Sigma(r)$ is a sharply declining transition by construction -- a direct consequence
of the background subtraction that defines a filament as a finite-extent sub-structure of the cloud -- and is therefore
intrinsically resolved in the data, with no internal length scale that the beam could fail to resolve. How accurately the entire
convolved profile, including its boundary region, is recovered by the fitting procedure is assessed in Sect.~\ref{convolved_models}
(Fig.~\ref{conv.accuracies}); that is the relevant diagnostic for convolution effects on $\Sigma(r)$ as a whole.

\subsubsection{Density slopes}

Accurate derivation of the surface and volume density slopes $\gamma$ and $\beta$ also requires high resolvedness
($\mathcal{R}\gtrsim 10$). At lower angular resolutions, both slopes become severely overestimated. For the shallowest models
($\beta_{\rm T}\approx 0.5$), the overestimation factor for $\beta$ reaches 20 in Fig.~\ref{conv.accuracies} (corresponding to
factors of approximately 30--50 for $\gamma$) when using the upper bound $\gamma_{\rm max\!}=9$ during fitting. These factors scale
proportionally with the upper bound, reaching 42 and 102 for $\gamma_{\rm max\!}=20$ and 50, respectively.

This dramatic overestimation occurs when the convolution beam becomes wide enough to transform the true $\Sigma(r)$ profile -- with
its increasingly steep decline to zero at $r\rightarrow R$ -- into a Gaussian-like profile that extends across and beyond the
boundary. At such low angular resolutions, the convolved profile becomes so severely deformed that it can no longer be represented
by the model in Eq.~(\ref{surface_density_fittingfun}). The fitting algorithm then converges to much steeper slopes $\gamma$ (and
$\beta$) that better match the Gaussian-like shape of the convolution beam rather than the true filament structure. Naturally, this
convolution effect becomes relatively less severe for intrinsically steeper profiles with $\beta_{\rm T}\gtrsim 1$.

\subsubsection{Implications for observations}

These results demonstrate that high angular resolution ($\mathcal{R}\gtrsim 10$) is required for accurate derivation of filament
properties. To illustrate the practical challenge: if the true half-maximum width of a filament in the nearby \object{Taurus}
region (distance of 140 pc) is $h_{\rm T}\approx 0.1$ pc, then the observational beam required to accurately derive this width from
fitting $\Sigma_{\rm O}(r)$ must be $O\lesssim 15${\arcsec}, with proportionally higher resolution needed for more distant regions.
Such resolutions are presently not available in far-infrared observations. The median widths of filaments observed in various
star-forming regions with \emph{Herschel} correspond to $\mathcal{R}\approx 3{-}8$ \citep[with $O\approx
18${\arcsec},][]{Arzoumanian_etal2019}, below the threshold for accurate parameter recovery. Although the \emph{hires} method
\citep{Men'shchikov2021a} can enhance the effective resolution of \emph{Herschel} surface densities by approximately 35\% (to
13.5{\arcsec}), this improvement is insufficient to overcome the fundamental resolution limitation.

The systematic errors and their trends shown in Fig.~\ref{conv.accuracies} arise because convolution with finite beams transforms
the model profiles differently depending on the filament width $h_{\rm T}$, slope $\beta_{\rm T}$, and extent $\xi_{\rm T}$. The
resulting distorted profiles deviate in different ways and to varying degrees from the idealized fitting model of
Eq.~(\ref{surface_density_fittingfun}), forcing the fitting algorithm to produce inaccurate parameter estimates.

Based on the model results, it would in principle be possible to define applicability domains of the fitting method in the
parameter space ($\mathcal{R}, \beta_{\rm T}, \xi_{\rm T}$) where reliable results can be expected. However, these domains would
depend on the true values $\beta_{\rm T}$ and $\xi_{\rm T}$ of the volume density distribution, which are precisely the unknown
quantities we seek to determine in real observations. This creates a fundamental limitation: without knowing the true parameters a
priori, we cannot reliably assess whether a given observation has sufficient resolution for accurate parameter recovery, nor can we
remove the convolution effects through deconvolution (see Appendix~\ref{deconvolution}). Therefore, the model-based applicability
domains cannot provide definitive guidance for observational studies, and results from filaments with $\mathcal{R}
\lesssim 10$ should be interpreted with considerable caution.

\subsection{The traditional fitting approach}
\label{traditional_approach}

The major issue with previous observational studies of filaments is that the Plummer function (Eq.~(\ref{plummerfun})) cannot
reproduce the decline of $\Sigma(r)$ profiles to zero as $r\rightarrow R$ at the outer boundary (Fig.~\ref{vsdensprofiles}),
which is expected when filaments are modeled as structures of finite extent.
Despite this practical limitation, previous studies successfully fit the Plummer function to hundreds of observed filaments
\citep[e.g.,][]{Arzoumanian_etal2011, Palmeirim_etal2013, Arzoumanian_etal2019}, which presents an apparent contradiction that
requires explanation. After background subtraction, the observed $\Sigma_{\rm O}(r)$ falls to zero at the apparent outer
extent of the filament -- a direct consequence of the subtraction itself, regardless of whether the embedded filament has a
sharp physical boundary inside the cloud. We use this apparent outer extent as the initial estimate $R_0$
(Sect.~\ref{initial_estimates}). Our description (Eq.~(\ref{surface_density_fittingfun})) then refines this to the model parameter
$R$ within $[0.9, 1.1]\,R_0$ and represents the shape of the decline of $\Sigma(r)$ near $r\to R$ through the $\epsilon$-term. The
modeling choice lies in this functional representation of the decline -- which the Plummer function (Eq.~(\ref{plummerfun})),
being an unbounded power law, lacks entirely -- not in the existence of the decline, which is created automatically by the
subtraction.

\subsubsection{Effects of profile averaging}

Previous studies applied Plummer fitting to median profiles computed along entire filaments \citep[e.g.,][]{Arzoumanian_etal2011,
Palmeirim_etal2013, Arzoumanian_etal2019}. For filaments with perfect cylindrical geometry, computing the median profile would
reduce random observational noise and background fluctuations, and the resulting smoother profile would accurately represent the
properties of the real filament. However, as discussed by \cite{Men'shchikov2023}, such idealized cylindrical structures do not
exist in reality. Observed filaments show significant variations in their profile properties (in terms of $H$, $\gamma$, and $R$)
along their crests. For long filaments, averaging these spatially varying profiles spreads out the steep decline of $\Sigma(r)$ to
zero at $r\rightarrow R$, effectively making it invisible in the median profile. Although such averaged surface density profiles
can be easily fit with the Plummer function, they do not accurately represent the true profiles of observed filaments with varying
properties along their crests.

In real observations the position of the apparent edge -- where the background-subtracted profile reaches zero -- is
itself uncertain because background subtraction is imperfect. With perfect subtraction, a filament with cylindrical geometry would
be fully described by Eq.~(\ref{surface_density_fittingfun}) for the appropriate $\epsilon$; in practice, under- or
over-subtraction of the background flattens or steepens the apparent decline near the edge, compounding the obscuring effect of
profile averaging discussed above and contributing to the apparent success of Plummer fitting on real observed profiles.

\subsubsection{Numerical test of median profile effects}

We devised a simple test to investigate quantitatively how using median profiles of spatially varying filaments affects the
resulting profile shapes and fitting results. The test was performed on model profiles with 10\% noise added but no background, so
that the effect of averaging is demonstrated in isolation from background subtraction uncertainty. A nonuniform filament with
varying profiles along its crest was simulated using the models of a long straight filament (Sect.~\ref{convolved_models}) with
$R_{\rm T}=0.5$ pc, $\xi_{\rm T}=$ \{1, 2\}, and $\beta_{\rm T}=$ \{1, 2, 3, 6\}. These ranges are representative of filaments
observed with \emph{Herschel}. The result is geometric in origin and robust to the specific choice of ranges; wider ranges yield
proportionally larger overestimation factors but the same systematic trend. The profiles were scaled radially multiple times using
random scaling factors (drawn from a normal distribution) in the interval $[0.7, 1.5]$ to create a distribution of 400 profiles
with boundary radii between $0.7R$ and $1.5R$ and correspondingly varying slopes $\beta$, widths $h$, and measured widths $H$. The
median profile computed from these varying profiles was then fitted with the Plummer function from Eq.~(\ref{plummerfun}).

\begin{figure}
\centering
\centerline{\resizebox{0.91\hsize}{!}{\includegraphics{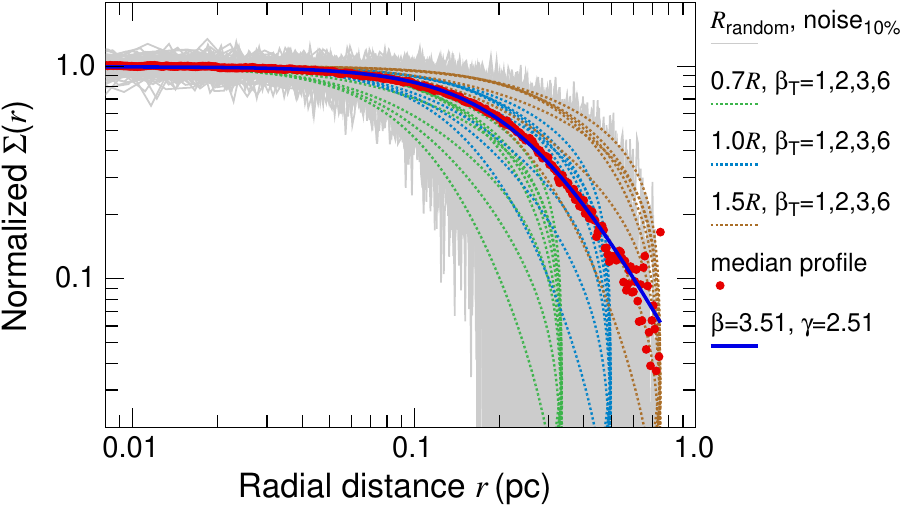}}}
\caption{
Simulation of the traditional approach of fitting median surface density profiles of entire filaments with the Plummer function.
Dotted blue, green, and brown curves show 8 representative original model profiles with $R=0.5$ pc, $\xi_{\rm T}=$ \{1, 2\}, and
$\beta_{\rm T}=$ \{1, 2, 3, 6\}, as well as the same profiles scaled to $0.7R$ and $1.5R$. Overlapping gray curves show 400
randomly scaled original profiles in the range $[0.7, 1.5]R$ with uniform noise at a 10\% level. Filled red circles display the
median profile points computed over the scaled profiles.
}
\label{fig:varying_median}
\end{figure}

Results are shown in Fig.~\ref{fig:varying_median}. The median profile, which differs significantly from the original individual
profiles, is well fitted by the Plummer function with a derived volume density slope $\beta=3.51$ and core width $2r_{\rm c}=0.52$
pc. However, this power-law median profile cannot reproduce the decline near the boundary present in the true surface density
profiles of the model filament at $r\gtrsim r_{\rm c}$. The derived slope and measured width $H=0.45$ pc from the Plummer fit are
overestimated relative to the median values of the individual model profiles ($\tilde{\beta}=2.5$ and $\tilde{H}=0.34$ pc) by
factors of 1.40 and 1.35, respectively. The median profile shows evidence of the boundary presence only at $r\gtrsim 0.65$ pc and,
unlike the true $\Sigma(r)$ profiles, begins a steeper decline very close to the fitted boundary $R=0.75$ pc. The latter is
overestimated by a factor of 1.50 relative to the true median boundary $\tilde{R}=0.50$ pc.

\subsubsection{Implications for mass and linear density estimates}

We also computed the linear densities $\Lambda$ and masses $M$ of the nonuniform and "median" filaments by integrating their true
simulated and fitted median profiles, respectively. The median values $\tilde{\Lambda}$ and $\tilde{M}$ computed from the
individual profiles of the nonuniform filament agreed within 1.1\% with the values derived from the fitted ``median'' filament
profile. However, the actual linear densities $\Lambda(l)$ of the simulated filament as a function of position $l$ along its crest
exhibited a wide range of values, varying by factors of 2.2 below and 1.75 above the median value. Although these specific factors
reflect the input scaling range, they illustrate the key point that this spatial variability is completely hidden by the median
profile, which agrees with the true median to within 1.1\%, creating a false impression of spatial uniformity. Such large spatial
variations along the filament are physically significant because linear densities are thought to determine fragmentation conditions
in observed filaments \citep[e.g.,][]{Arzoumanian_etal2019}.

\begin{figure}[ht!]
\centerline{\resizebox{0.77\hsize}{!}{\includegraphics{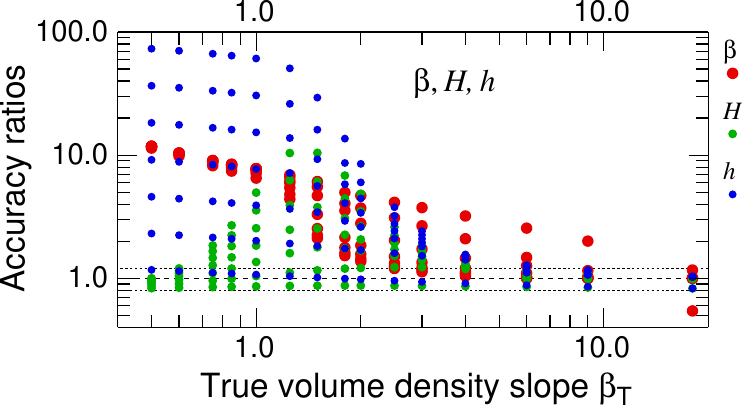}}}
\caption{
Accuracies of traditional fitting with the Plummer function of Eq.~(\ref{plummerfun}) for numerical model profiles (cf.
Fig.~\ref{modelacc}). The accuracy ratios $\beta\beta_{\rm T}^{-1}$ and $HH_{\rm T}^{-1}$ are shown with filled circles for surface
density profiles without noise. Vertical spread of the points is caused by different $\xi_{\rm T}$ values. \chgn{Also shown is the
accuracy $hh_{\rm T}^{-1}$ of the volume density width that the traditional approach infers using its implicit assumption $h=H$; 
this ratio reveals the systematic error incurred by that assumption, which grows very large for shallow, extended profiles (cf. 
Fig.~\ref{relationships}).} The corresponding plot with 10\% noise added is visually indistinguishable from this one. The fitting 
(in logarithmic space) was constrained with an upper bound of $\beta_{\rm max\!}=21$.
} 
\label{modelacc_plummer}
\end{figure}

\begin{figure*}[ht!]
\centerline{\resizebox{0.3090\hsize}{!}{\includegraphics{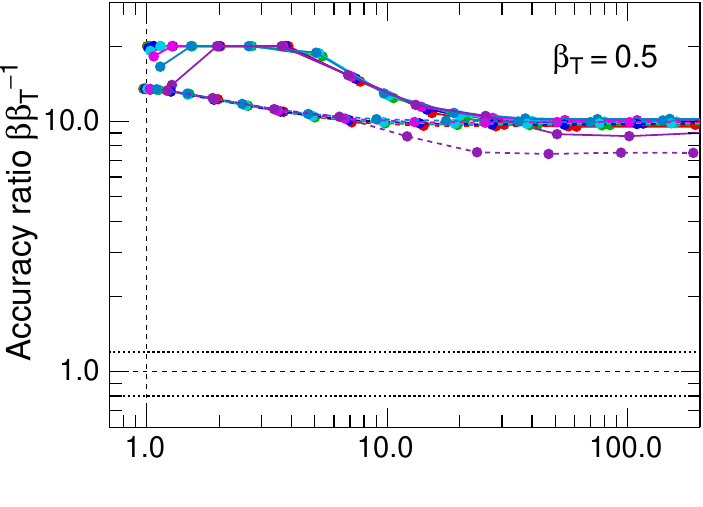}}
            \resizebox{0.2614\hsize}{!}{\includegraphics{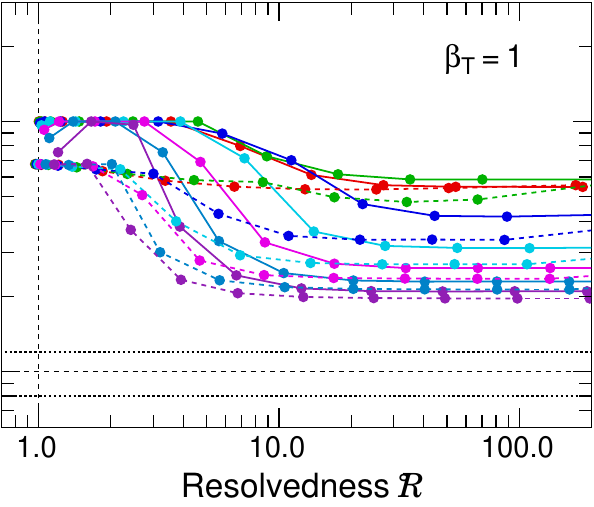}}
            \resizebox{0.3053\hsize}{!}{\includegraphics{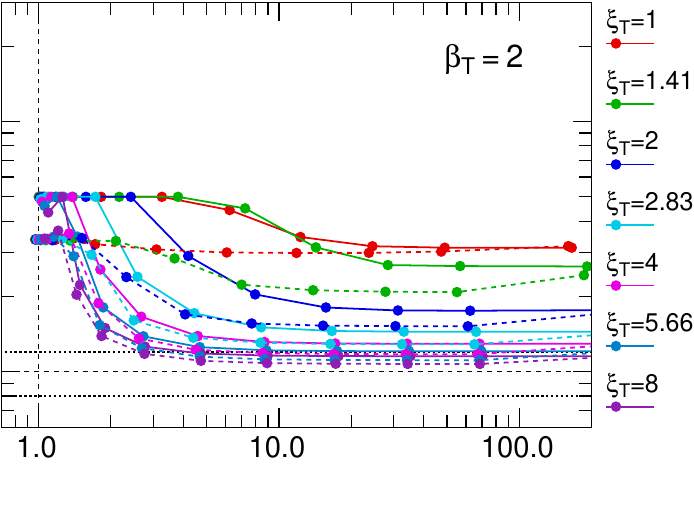}}}
\caption{
Accuracies of traditional fitting with the Plummer function of Eq.~(\ref{plummerfun}) for surface density profiles of a model
filament convolved with increasingly larger beams (cf. Fig.~\ref{conv.accuracies}). The accuracy ratios $\beta\beta_{\rm T}^{-1}$
are shown with filled circles connected by solid lines for different $\beta_{\rm T}$ and $\xi_{\rm T}$. The accuracies for modified
profiles truncated at the $0.1\Sigma_{\rm C}$ level and altered by 10\% noise are connected by dashed lines.
}
\label{conv_plummer_accuracies}
\end{figure*}

These results demonstrate that while median profiles may provide reasonable estimates of average filament properties, they obscure
potentially important spatial variations. A more appropriate approach for detailed characterization of spatially varying profiles
and linear densities is to divide observed filaments into short segments and analyze the profiles over each segment length
individually \citep{Zhang_etal2024,ZhangMenshchikovLi2026}. This segmented approach preserves information about spatial
variations while still allowing for noise reduction within each segment.

\subsection{Performance of the Plummer fitting function}
\label{comparison_old_method}

To evaluate the performance of the traditional Plummer fitting function, we repeated the same validation tests on model and
observed profiles that were described in Sect.~\ref{validation}. The profiles were fitted with the three free parameters of
Eq.~(\ref{plummerfun}): the crest surface density $\Sigma_{\rm C}$, the core radius $r_{\rm c}$, and the slope $\beta$.

\subsubsection{Non-convolved models}

Results for non-convolved model profiles (cf. Sect.~\ref{pure_models}) are displayed in Fig.~\ref{modelacc_plummer} in terms of the
accuracies of the derived slopes $\beta$ and widths $H$. As throughout the paper, $H$ denotes the measured half-maximum width of
$\Sigma(r)$, not the Plummer core radius $r_{\rm c}$; the accuracy ratio $HH_{\rm T}^{-1}$ compares $H$ from the Plummer fit to the
true model value $H_{\rm T}$. The traditional Plummer model (Eq.~(\ref{plummerfun})) provides no volume density width $h$; it
measures $H$ and implicitly assumes $h\approx H$. \chgn{Observers are ultimately interested in volume density widths $h$, therefore
Fig.~\ref{modelacc_plummer} also shows their accuracy $hh_{\rm T}^{-1}\equiv Hh_{\rm T}^{-1}$ under the traditional assumption.
This ratio incorporates the intrinsic mismatch between $H$ and $h$ quantified in Fig.~\ref{relationships}; it is substantially
worse than the accuracy of $H$ for shallow, extended filaments, where $H$ increasingly overestimates $h$ (by almost two orders of
magnitude for large $\xi$).} The addition of noise to the profiles does not lead to substantial degradation of the derived
parameters, indicating that the Plummer fitting is robust to typical noise levels.

However, very significant systematic errors are evident. The slopes $\beta$ become increasingly overestimated toward shallower
profiles, roughly in inverse proportion to the true slope $\beta_{\rm T}$, reaching order-of-magnitude errors at $\beta_{\rm T}
\lesssim 1$. A substantial spread in the slope accuracies for profiles of different extents ($\xi_{\rm T} =$ \{1, 2, 4, 8, 16, 32,
64\}) exists for $1 \lesssim \beta_{\rm T} \lesssim 10$. Profiles with larger $\xi_{\rm T}$ achieve better accuracies because they
have broader power-law segments that are better approximated by the Plummer function (Fig.~\ref{vsdensprofiles}).
For very shallow profiles ($\beta_{\rm T}\lesssim 0.7$), in contrast, the spread across $\xi_{\rm T}$ is small. The Plummer
function, having no outer boundary, cannot reproduce the steep decline of the model profile near $r=R$, so the fit is driven
primarily by the overall profile shape and depends only weakly on the relative position of the boundary set by
$\xi_{\rm T}$.

The half-maximum widths $H$ are reasonably well determined for the shallowest ($\beta_{\rm T} \lesssim 0.7$) and steepest
($\beta_{\rm T} \gtrsim 3$) profiles, whereas for intermediate slopes ($0.7\lesssim\beta_{\rm T}\lesssim 3$) they become greatly
overestimated (Fig.~\ref{modelacc_plummer}). This occurs because Eq.~(\ref{plummerfun}) cannot reproduce the steep decline of
$\Sigma(r)$ to zero near the outer boundary in finite-extent filament models, and this limitation is most severe for profiles with
intermediate slopes and large extents $\xi_{\rm T}$.

\subsubsection{Convolved models}

Results for models convolved to various angular resolutions (cf. Sect.~\ref{convolved_models}) are presented in
Fig.~\ref{conv_plummer_accuracies}. For the shallowest profiles ($\beta_{\rm T}=0.5$), the derived slope $\beta$ is overestimated
by more than an order of magnitude across the entire range of resolvedness values. These large systematic errors progressively
decrease for steeper profiles, until the derived slope appears accurate within 20\% for $\beta_{\rm T\!}=9$. However, this
seemingly accurate result is an artifact caused by the upper bound $\beta_{\rm max\!}=10$ used in the constrained fitting. With a
twice higher bound ($\beta_{\rm max\!}=20$), the slope becomes overestimated by a factor of two even for $\beta_{\rm T\!}=9$, and
the overestimation factors for shallower slopes increase proportionally. The accuracies do become progressively better for both
original and modified profiles with larger extents ($\xi_{\rm T}\gtrsim 3$), as expected from the non-convolved results.

\subsubsection{Application to observed California filaments}

\begin{figure}
\centering
\centerline{\resizebox{0.9\hsize}{!}{\includegraphics{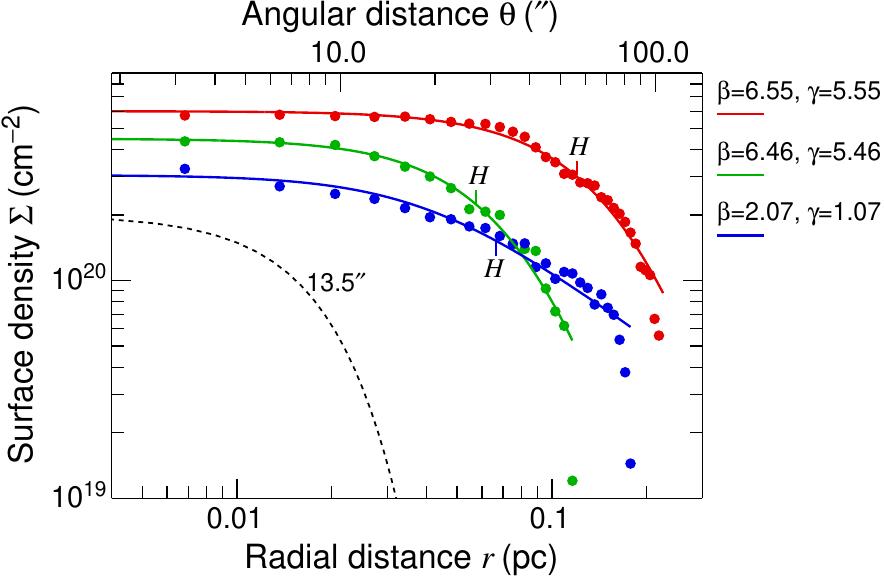}}}
\caption{
Application of the traditional fitting method of Eq.~(\ref{plummerfun}) to the observed filament profiles shown in
Fig.~\ref{fig:california_profiles}. The derived $\beta$ and $\gamma=\beta-1$ are specified, and the half-maximum widths $H$ are
indicated next to the curves at the radial positions of their half-widths. 
The Plummer fitting implicitly assumes $h\approx H$.
}
\label{fig:california_plummer}
\end{figure}

Results of traditional Plummer fitting for three observed profiles of filaments in the \object{California} star-forming region (cf.
Sect.~\ref{applied_to_California}) are shown in Fig.~\ref{fig:california_plummer}. The data points are not fitted well by the
Plummer function, with particularly poor agreement near the boundaries where the observed profiles decline steeply to zero. The
goodness-of-fit values $\textsc{R}^2=$ \{0.962, 0.987, 0.981\} for the steep, intermediate, and shallow profiles are lower than
those of our new method (\{0.987, 0.993, 0.991\}), quantifying the poorer quality of the Plummer fits. The most problematic
finding is that the derived slopes $\beta =$ \{6.55, 6.46, 2.07\} (corresponding to $\gamma = \beta - 1 =$ \{5.55,
5.46, 1.07\}) are much steeper than those derived using our method ($\beta =$ \{3.47, 1.57, 0.74\},
Fig.~\ref{fig:california_profiles}). The formal uncertainties on the Plummer slopes are
$\sigma_\gamma=$ \{1.91, 2.05, 0.140\} and on the dense core radii $\sigma_{r_{\rm c}}=$ \{0.046, 0.024, 0.006\} pc
for the steep, intermediate, and shallow profiles, respectively. The large $\sigma_\gamma$ for
the two steep profiles reflects the same difficulty in constraining steep slopes, as discussed in
Sect.~\ref{quality}; notably, the new method yields considerably smaller uncertainties
$\sigma_\gamma=$ \{1.20, 0.107, 0.029\} for the same profiles (Sect.~\ref{applied_to_California}).
Although the true slopes of these profiles are not known, the poor quality of the fits
combined with the systematic biases revealed by our model tests (described above) strongly suggest that the Plummer-derived slopes
are unreliable.

In contrast, the half-maximum widths derived from Plummer fitting ($H =$ \{0.265, 0.105, 0.136\} pc) are practically the same as
those obtained with our method ($H =$ \{0.258, 0.109, 0.137\} pc, Sect.~\ref{applied_to_California}). This similarity is not
surprising, as the half-maximum points located in the middle portion of the profiles are usually fitted fairly well by any
reasonable function, unless the fit is completely inappropriate. However, this agreement in $H$ does not validate the Plummer
function, as the derived slopes and physical interpretation remain highly questionable.

\subsection{Deconvolution of surface density profiles?}
\label{surfdens_deconv}

Insufficient angular resolution strongly affects the parameters of filaments (and other sources) derived from fitting their surface
densities. For this reason, a popular approach has been to deconvolve the measured half-maximum widths $H$ of the observed
$\Sigma_{\rm O}(r)$ profiles to estimate the ``true'' widths $\breve{H}$ of filaments unaffected by finite observational
resolution. However, the simple and convenient Gaussian deconvolution using Eq.~(\ref{gauss_deconv}) is valid only for structures
with steep Gaussian-like profiles and is generally not applicable to ordinary power-law profiles due to large systematic errors
\citep{Men'shchikov2023}.

\subsubsection{Limitations of model-based deconvolution}

In principle, the accuracy curves from Fig.~\ref{conv.accuracies} can be inverted and viewed as deconvolution functions for all
parameters ($h$, $\beta$, $w$, $\gamma$, $\xi$, etc.). However, they are impractical for real observations because of the large
systematic errors in the derived $\beta$ and $\xi$ at low resolvedness ($\mathcal{R}\lesssim 10$). Critically, the results in
Fig.~\ref{conv.accuracies} are based on the known true values $\beta_{\rm T}$ and $\xi_{\rm T}$, which are precisely the unknown
quantities we seek to determine from observations. Therefore, these curves primarily serve to demonstrate the magnitude of
systematic errors that convolution can induce on parameters derived from fitting, rather than providing a practical deconvolution
method.

\subsubsection{Deconvolution formula for ``infinite'' profiles}

Having investigated the effects of angular resolution on the measured widths $H$ of filament surface densities from
Eq.~(\ref{surface_density_fittingfun}) through a series of convolutions of model images with Gaussian beams, we derived a new
deconvolution formula (Appendix~\ref{infinite}). The more general Eq.~(\ref{fil_deconv}) is valid for ``infinite'' power-law
profiles $\Sigma(r)$ with accurately known slopes $\gamma$, as well as for Gaussian profiles. In practical terms, infinite profiles
(with $R\rightarrow \infty$) can be defined as those with extremely wide footprints ($RH^{-1}\gg 1$). 

Although Eq.~(\ref{fil_deconv}) is applicable for deconvolving $H$ of such extended power-law structures, the accuracy of width
deconvolution in astrophysical applications is likely to be very poor for two reasons. First, large uncertainties in the derived
and subtracted backgrounds lead to deformed profiles \citep{Men'shchikov2023}. Second, the power-law slopes $\gamma$ measured from
observed profiles are likely inaccurate because the observed $\Sigma_{\rm O}(r)$ profiles deviate significantly from pure power
laws (Figs.~\ref{vsdensprofiles}, \ref{fig:california_profiles}), particularly near their boundaries.

\subsubsection{The problem: volume density vs. surface density}

Most importantly, the entire concept of size deconvolution aims to derive the physical widths $h$ of the volume density
distributions $\rho(r)$ rather than the widths $H$ of the projected (integrated) surface density profiles $\Sigma_{\rm O}(r)$. As
demonstrated in Figs.~\ref{relationships} and \ref{fig:california_profiles}, these two widths can differ quite significantly,
particularly for shallow slopes and extended profiles. Our numerical results for $H$ and $h$ in Appendix~\ref{finite}
(Fig.~\ref{deconvolfin}) illustrate the general deconvolution functions for power-law profiles with arbitrary footprints. We did
not derive analytical approximations for these numerical functions because their practical applicability would require accurate a
priori knowledge of both the slope ($\beta\approx \beta_{\rm T}$) and extent ($\xi\approx \xi_{\rm T}$). However, from
Sect.~\ref{convolved_models} we know that these parameters, when determined from fitting the observed $\Sigma_{\rm O}(r)$ profiles
of power-law structures at low $\mathcal{R}$, have very large systematic errors (Fig.~\ref{conv.accuracies}).

This creates a fundamental circular problem: accurate deconvolution requires knowing the true parameters $\beta_{\rm T}$ and
$\xi_{\rm T}$, but these are precisely what we are trying to measure from the convolved observations. The root cause of this
limitation is that poor angular resolution leads to a severe loss of information about the observed structures and their true
profiles, making it impossible to reliably recover the intrinsic parameters through deconvolution alone. The only robust solution
to this problem is to obtain observations with sufficiently high angular resolution ($\mathcal{R} \gtrsim 10$) in the first place.


\section{Conclusions}
\label{conclusions}

We have developed a new method for deriving the physical properties of filamentary structures from observed surface density
profiles. The method addresses fundamental limitations of the traditional Plummer function approach by explicitly accounting for
the finite nature of filaments. 

We derived empirical analytical relationships that connect the parameters of surface density profiles $\Sigma(r)$ to those of
volume density profiles $\rho(r)$, accounting for the finite cylindrical geometry of filaments. The difference $\delta = \beta -
\gamma$ between volume and surface density slopes is not constant but varies systematically with the slope and extent of the
filament, progressively decreasing below unity for shallow and compact profiles. This contrasts with the fixed value $\delta = 1$
implicitly assumed in previous studies. The half-maximum widths $h$ and $H$ of the volume and surface density profiles can differ
substantially, particularly for extended filaments with shallow slopes. Traditional approaches that assume $h \approx H$ can lead
to order-of-magnitude errors in the derived physical widths.

For non-convolved model profiles, the method achieves high accuracy across a wide range of slopes and extents, and remains robust
to moderate noise levels. The most significant limitation is the strong dependence on angular resolution -- a fundamental
limitation for all such methods. Accurate parameter recovery requires very high resolvedness for shallow slopes and moderate
resolvedness for steeper slopes. At lower resolvedness, systematic errors become severe, with slopes being particularly affected.
Contrary to the traditional assumption, filaments with shallow slopes remain effectively unresolved even when their measured width
$H$ is much larger than the beam size. This has important implications for the interpretation of existing observations,
particularly from \emph{Herschel}.

The Plummer function cannot reproduce the steep decline of $\Sigma(r)$ to zero at the outer boundary expected for filaments modeled
as structures of finite extent, and yields systematically overestimated slopes, particularly at low $\mathcal{R}$. Computing median
profiles along entire filaments obscures boundary effects and loses information about spatial variations in filament properties.

Accurate deconvolution of observed profiles requires a priori knowledge of the true parameters, creating a fundamental circular
problem. The only robust solution is to obtain observations with sufficiently high angular resolution. We recommend prioritizing
the highest possible angular resolution, interpreting parameters from filaments with $\mathcal{R}\lesssim 10$ with caution,
dividing long filaments into short segments to preserve spatial variations, and reporting the resolvedness along with derived
parameters.

Our results highlight the critical need for higher angular resolution observations to enable reliable characterization of the
physical structure of filamentary clouds in star-forming regions. The systematic biases we have identified suggest that some
previously reported filament properties may require reinterpretation.

\begin{acknowledgements}
The \emph{plot} utility and \emph{ps12d} library, used to draw figures directly in the PostScript language, were written by A.M.
using the \emph{psplot} library (by Kevin E. Kohler), developed at Nova Southeastern University Oceanographic Center (USA), and the
plotting subroutines from the MHD code \emph{azeus} \citep{Ramsey2012}, developed by David Clarke and A.M. at Saint Mary's
University (Canada). We thank the referee, V.~Ossenkopf-Okada, for the careful and constructive reports, which have improved 
the clarity and rigor of this paper.
\end{acknowledgements}

\bibliographystyle{aa} 
\bibliography{svdensities}

\begin{appendix}
\onecolumn
\section{Analytical relationships}
\label{app:relationships}

\subsection{Auxiliary parameters}
\label{app:auxiliary}

This appendix collects the analytical relationships that are not needed to interpret the physical parameters $\beta$ and $h$
(Eqs.~(\ref{betaformula}) and (\ref{hHformula})), but that are required to evaluate the fitting function
Eq.~(\ref{surface_density_fittingfun}) at every iteration of the fit and, therefore, to reproduce the algorithm 
(Sect.~\ref{method}). They were obtained by the same empirical procedure, described in Sect.~\ref{fittingfun}.

\begin{figure*}[ht!]
\centerline{\resizebox{0.3449\hsize}{!}{\includegraphics{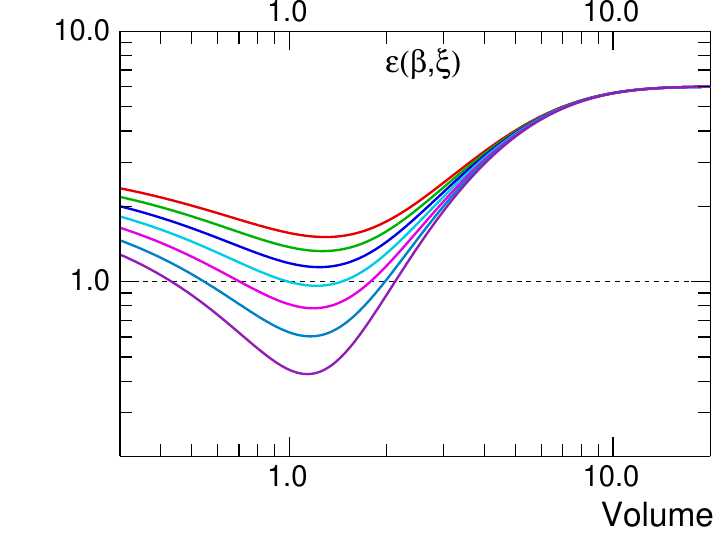}}
            \resizebox{0.3600\hsize}{!}{\includegraphics{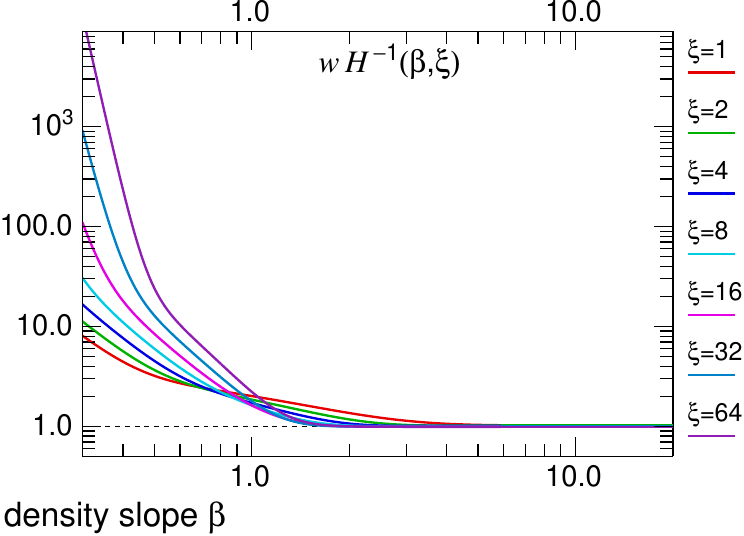}}}
\caption{
\chgn{Analytical dependencies of the auxiliary parameters on $\beta$ for selected values of $\xi$ (Eqs.~(\ref{epsilonformula}) and 
(\ref{wHformula})): the boundary exponent $\epsilon$ (\emph{left}) and the ratio of the intrinsic and measured half-maximum widths 
$wH^{-1}$ (\emph{right}). These complement the dependencies of the slope $\gamma$ and the width ratio $hH^{-1}$ shown in 
Fig.~\ref{relationships}.}
}
\label{relationships_aux}
\end{figure*}

The boundary exponent $\epsilon$ in Eq.~(\ref{surface_density_fittingfun}) can be approximated as a function of $\xi$ and $\beta$:
\begin{equation}
\epsilon = 40.115 \left(1 + \exp\left(-0.3782 \left(\beta + 2.542\right)\right) \xi^{0.012}\right)^{-1} + 8.21 \exp\left(-0.7497 \beta \xi^{0.0475}\right) \xi^{-0.015} - 34.104.
\label{epsilonformula}
\end{equation}
which exhibits a minimum near $\beta \approx 1.2$ (\chgn{Fig.~\ref{relationships_aux}}). This indicates that $\Sigma(r)$ profiles with such
slopes decline to zero at $r \rightarrow R$ more gradually than profiles with significantly shallower or steeper slopes. For $\beta
\gtrsim 4$, the exponent $\epsilon$ becomes essentially independent of the extent $\xi$.

The intrinsic half-maximum width $w$ of the surface density profile $\Sigma(r)$ can be approximated as a function of the
measured half-maximum width $H$ of the observed profile $\Sigma_{\rm O}(r)$, along with $\xi$ and $\beta$:
\begin{equation}
w = H \left(235.7\exp\left(-20\beta \xi^{-0.2}\right) + 0.00005\xi^{0.5}\beta^{-6\xi^{0.21\!}\!}+2.878\exp\left(-1.069\beta \xi^{0.22\beta}\right) + 113.0\exp\left(-10.88\beta \xi^{-0.2}\right)+ 1.022\xi^{-0.0077} \right).
\label{wHformula}
\end{equation}
For the shallowest profiles, the intrinsic width $w$ is significantly larger than the measured width $H$, whereas for $\beta
\gtrsim 3$, the two widths become essentially identical (\chgn{Fig.~\ref{relationships_aux}}). The divergence between $w$ and $H$ at lower
$\beta$ and larger $\xi$ (illustrated in Fig.~\ref{vsdensprofiles}) arises from the geometrical boundary term (the square-root 
factor with exponent $\epsilon$) in Eq.~(\ref{surface_density_fittingfun}).

Finally, the coefficients $E$, $F$, $G$, $S$, and $Z$ of Eq.~(\ref{hHformula}) are the following functions of $\xi$:
\begin{align}
\begin{split}
E &= 0.77149\left(1-\exp\left(-\left(\left(\xi-0.52709\right)/0.7156\right)^{-0.9095}\right)\right)+0.0026857, \\
F &= -0.31586\xi^{-1.9388\!}+0.57344\xi^{-0.96778\!}+6.5472\xi^{0.14471} - 5.1551, \\
G &= 1.0811\xi^{-0.62466\!}-1.4375\xi^{-1.0203\!}+4.0626\xi^{0.00955} - 3.1165, \\
S &= 0.034613 \exp\left(-0.014394\xi\right) +0.036328\exp\left(-0.27696\xi\right)+2.1537\exp\left(-5.104\xi\right) + 0.94275, \\
Z &= 0.26355 \exp\left(-0.059635\xi\right)+8.8497\exp\left(-4.824\xi\right) -645.57\exp\left(-12.634\xi\right) + 0.24014.
\label{coeffshH}
\end{split}
\end{align}

\clearpage
\subsection{Intrinsic accuracies}
\label{app:accuracy}

\begin{figure*}[ht!]
\centerline{\resizebox{0.3236\hsize}{!}{\includegraphics{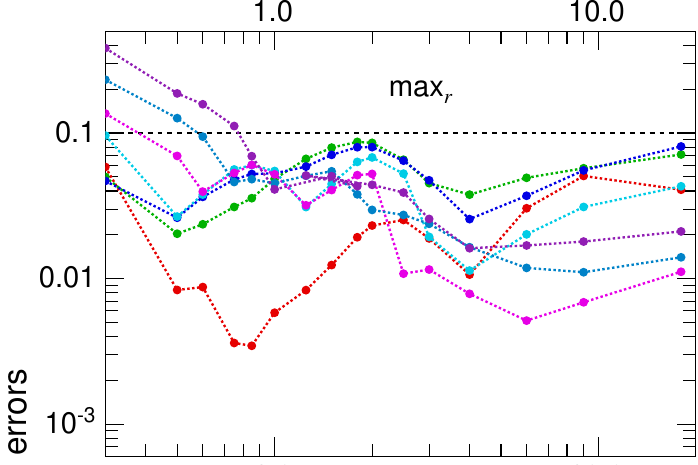}}
            \resizebox{0.3081\hsize}{!}{\includegraphics{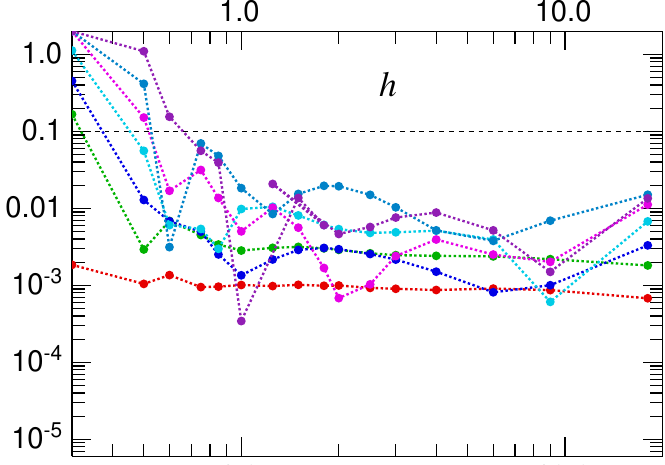}}
            \resizebox{0.3120\hsize}{!}{\includegraphics{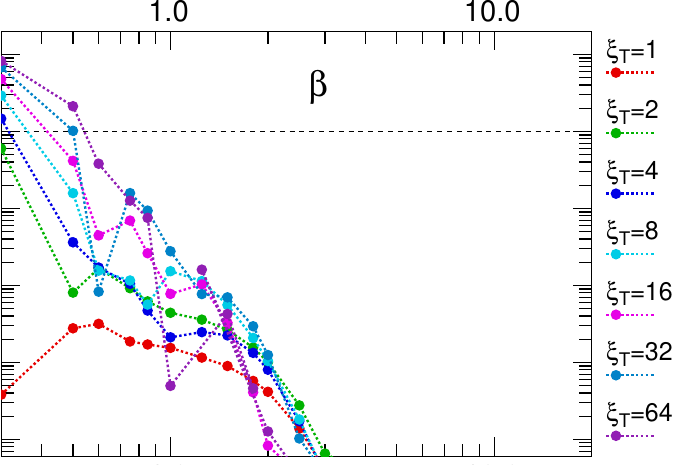}}}
\centerline{\resizebox{0.3236\hsize}{!}{\includegraphics{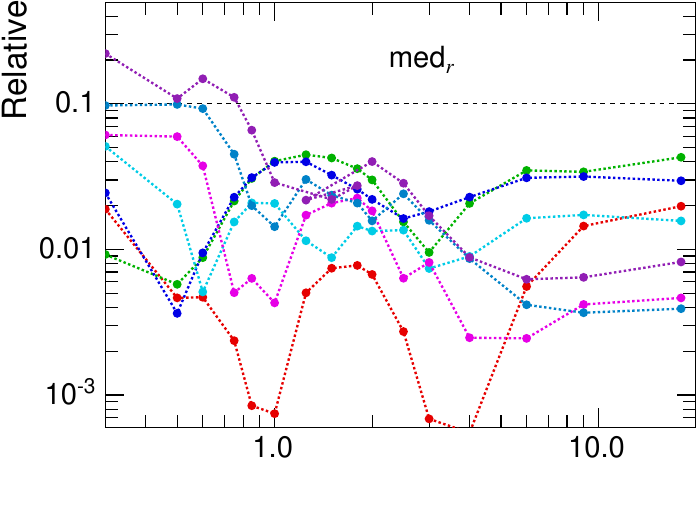}}
            \resizebox{0.3081\hsize}{!}{\includegraphics{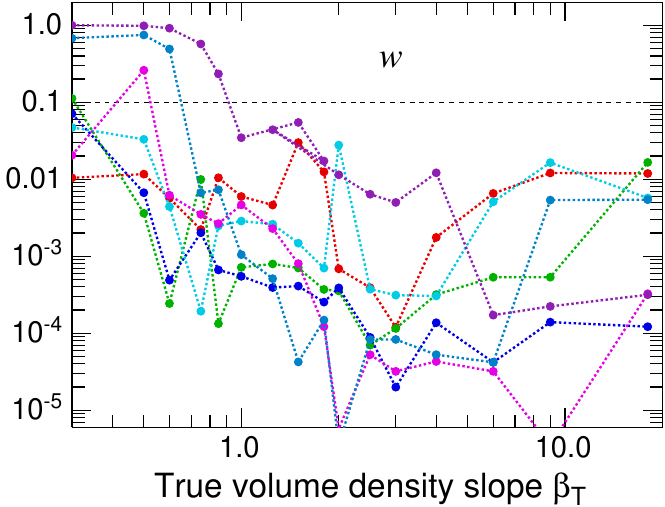}}
            \resizebox{0.3120\hsize}{!}{\includegraphics{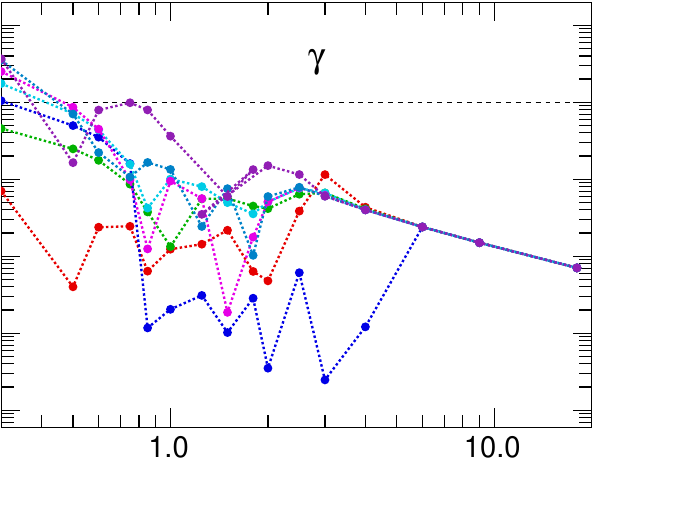}}}
\caption{
Accuracy of the analytical profiles and their parameters, as functions of $\beta_{\rm T}$ and $\xi_{\rm T}$. The \emph{left} panels
show the maximum and median relative deviations of the analytical surface density profiles from the numerical models. Relative
errors in the main parameters $h$, $\beta$, $w$, and $\gamma$ (\emph{middle} and \emph{right}), derived from
Eqs.~(\ref{betaformula}), (\ref{hHformula}), and (\ref{wHformula}), are defined with respect to the true values $h_{\rm T}$ and
$\beta_{\rm T}$ used to compute the numerical models and the optimal values $w_{\rm T}$ and $\gamma_{\rm T}$ used to obtain the
analytical approximations. The dashed horizontal lines mark the 10\% error level.
} 
\label{relerrors}
\end{figure*}

Figure~\ref{relerrors} documents the intrinsic accuracy of the relationships collected above and of Eqs.~(\ref{betaformula}) and
(\ref{hHformula}), for the numerical models of Sect.~\ref{validation}. The maximum and median relative deviations of the analytical
profiles from the numerical ones are quite low across nearly the entire parameter space. The largest deviations occur only for very
shallow slopes ($\beta_{\rm T} < 0.7$) and extended profiles ($\xi_{\rm T}\gtrsim 10$), and are local to radii near the outer
boundary, where the surface density declines extremely steeply to zero as $L(r)\rightarrow 0$. The relative errors in the derived
parameters $h$, $\beta$, $w$, and $\gamma$ are also low across nearly the entire parameter space, with typical errors below a few
percent. The scatter visible between adjacent points in these panels reflects the fact that the values of $w$, $\gamma$, and
$\epsilon$ were determined independently for each model by visual matching of Eq.~(\ref{surface_density_fittingfun}) to
$\Sigma_{\rm N}(r)$ (Sect.~\ref{fittingfun}); each point therefore represents an independent visual fit. The analytical curves
are smooth approximations through this discrete set of values, and the residual scatter is their irreducible deviation from the
per-model results. The larger deviations for $\beta_{\rm T}\lesssim 0.7$ and $\xi_{\rm T}\gtrsim 10$ arise because the visual
matching is intrinsically less precise there (Sect.~\ref{fittingfun}), as all three quantities become increasingly sensitive
functions of $\beta$ and $\xi$ in this regime. Such extreme parameter combinations are unlikely to occur in practice. The maximum
relative errors in the two parameters of direct physical interest, $\beta$ and $h$, are summarized in Table~\ref{accuracytable}
(Sect.~\ref{accuracy.analytical}).

\begin{figure*}[ht!]
\centerline{\resizebox{0.3300\hsize}{!}{\includegraphics{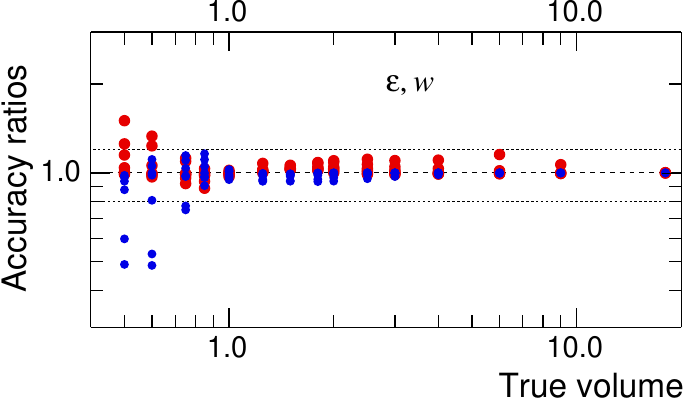}}
            \resizebox{0.3072\hsize}{!}{\includegraphics{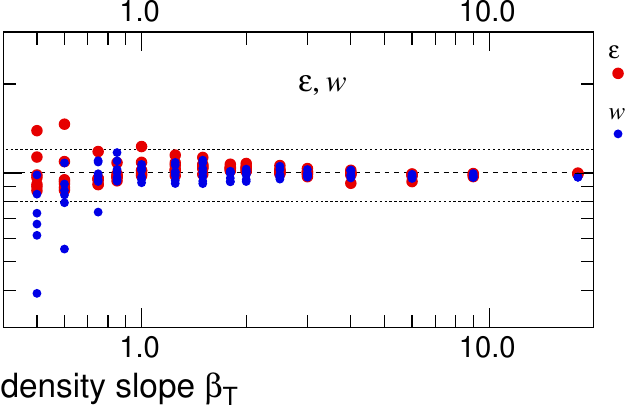}}}
\caption{
\chgn{Accuracies of the auxiliary parameters $\epsilon$ and $w$ derived from fitting the numerical model profiles 
(Eqs.~(\ref{epsilonformula}) and (\ref{wHformula})). The accuracy ratios with respect to the 
true model values are shown as functions of the true $\beta_{\rm T}$ for fitted surface density profiles without noise (\emph{left}) 
and with 10\% noise added (\emph{right}). Filled circles represent the parameters indicated at the edge of the right panel; their
vertical spread reflects variations due to different $\xi_{\rm T}$ values. Horizontal dashed lines indicate perfect accuracy, while
dotted lines mark $\pm 20\%$ error bounds. This figure complements the dependencies of the main parameters shown in 
Fig.~\ref{modelacc}.}
}
\label{modelacc_aux}
\end{figure*}

\chgn{Figure~\ref{modelacc_aux} shows the accuracies of the auxiliary parameters $\epsilon$ and $w$ obtained by fitting the
numerical model profiles, complementing the physical and fitted parameters presented in Fig.~\ref{modelacc}. These two quantities
enter the surface density model of Eq.~(\ref{surface_density_fittingfun}) but are usually not needed by an observer, who looks for
only $\beta$ and $h$; their accuracies are included here for completeness.}

\twocolumn
\section{Width deconvolution}
\label{deconvolution}

As discussed by \cite{Men'shchikov2023}, there are large uncertainties in deriving the ``true'' widths of power-law surface density
profiles of filaments (or sources) when applying the widely used Gaussian \citep[``naive'',][]{Andre_etal2022} deconvolution,
\begin{equation}
\breve{H} = H \left(1 - \mathcal{R}^{-2}\right)^{1/2},
\label{gauss_deconv}
\end{equation}
where $\breve{H}$ is the deconvolved half-maximum width, $\mathcal{R}>1$ the resolvedness, and $O$ the half-maximum width of the
observational beam represented by a Gaussian. As a consequence of its very large inaccuracies, Eq.~(\ref{gauss_deconv}) is not
applicable to filaments (or sources) with power-law density profiles.

\subsection{``Infinite'' (very extended) filaments}
\label{infinite}

\begin{figure}
\centerline{\resizebox{0.8\hsize}{!}{\includegraphics{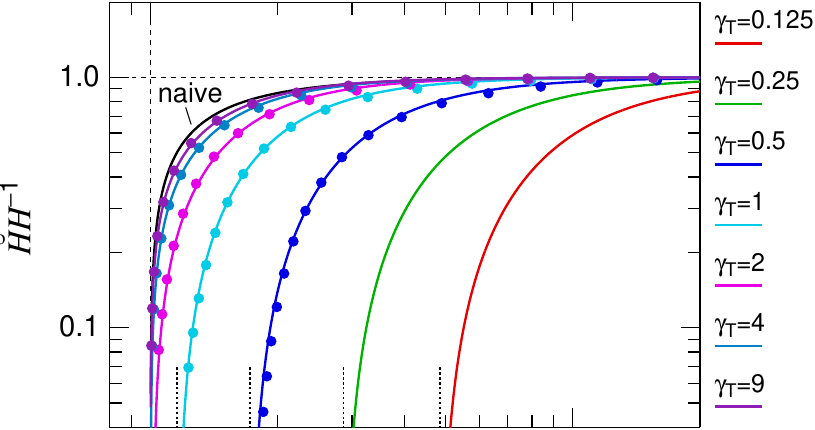}}}
\centerline{\resizebox{0.8\hsize}{!}{\includegraphics{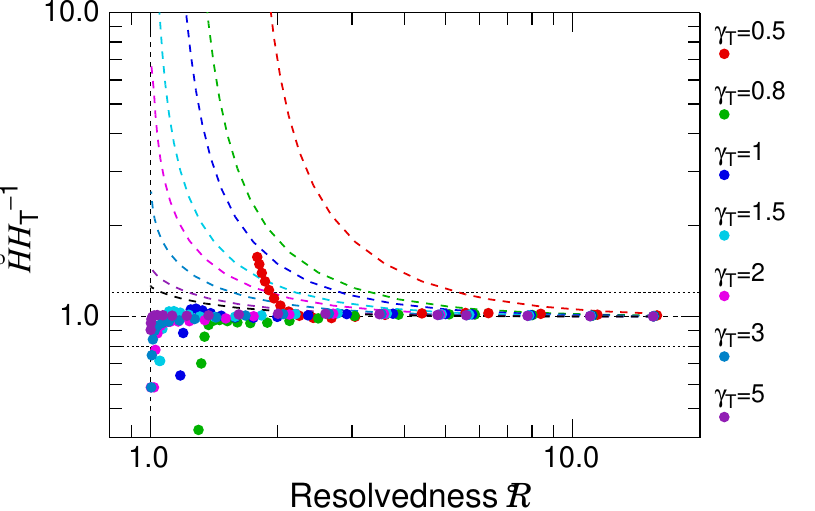}}}
\caption{
Deconvolution function (\emph{top}) for ``infinite'' power-law filaments (Sect.~\ref{infinite}) and its accuracies
(\emph{bottom}). The filled circles and solid curves show the deconvolution function $\breve{H}H^{-1}$
from convolved numerical models and Eq.~(\ref{fil_deconv}), respectively. The accuracy ratios $\breve{H}H_{\rm T}^{-1}$
are shown with filled circles, whereas for the naive deconvolution from Eq.~(\ref{gauss_deconv}) they are traced by dashed
curves of the same colors. The dashed horizontal line indicates perfect accuracy and two dotted lines show errors of
$\pm 20\%$. For reference, the black solid curve displays the function from Eq.~(\ref{gauss_deconv}). The dashed lines
indicate limiting values on both axes. For shallower slopes $\gamma$, the lower bounds for resolvedness progressively shift to
larger values $\mathcal{R}_{\rm min\!} = B^{1/A} =$ \{4.84, 2.86, 1.72, 1.16\} for
$\gamma_{\rm T}=$ \{0.125, 0.25, 0.5, 1\}, respectively; for steeper slopes $\mathcal{R}_{\rm min}\rightarrow 1$.
These bounds are indicated by dotted vertical lines in the top panel and serve as a
reference for both panels. The bottom panel shows slopes $\gamma_{\rm T}\geq 0.5$, so
its leftmost reference line is at $\mathcal{R}_{\rm min}=1.72$.
} 
\label{deconvolinf}
\end{figure}

We investigated the effects of angular resolution $O$ on the measured widths $H$ of filament surface densities from
Eq.~(\ref{surface_density_fittingfun}) in a series of convolutions of model images with Gaussian beams in the range 9.5 to
1730\arcsec. The model filaments were extremely wide ($RH^{-1}\gtrsim 60$) and their footprints occupied the entire image. On the basis
of these ``infinite'' models, we derived an empirical deconvolution formula for extremely wide filaments,
\begin{equation}
\breve{H} = H \left(\frac{\mathcal{R}^{A} - B} {\mathcal{R}^{A} + B}\right)^{C},
\label{fil_deconv}
\end{equation}
where resolvedness is limited to $\mathcal{R} > B^{1/A}$ and the coefficients $A$, $B$, and $C$ are functions of the slope $\gamma$
(assumed to be accurately known),
\begin{align}
\begin{split}
A &= 2.56-0.5\left(\left(\gamma+0.43\right)^{3.8}+0.847\right)^{-1}, \\
B &= 1+\left(0.55\left(\gamma+0.21\right)^{5} + 0.9\gamma^{1.5}\right)^{-1}, \\
C &= 0.477 + 2.77\left(0.007\left(\gamma+2.14\right)^{4.5} +3.9\right)^{-1}.
\end{split}
\label{auxiliaryfun}
\end{align}
The condition of very wide footprints ($RO^{-1\!}\gtrsim 5$) implies that convolution of the densest inner core of a structure is not
affected by boundary presence for any beam width $O$. The new deconvolution function and its accuracies are shown in
Fig.~\ref{deconvolinf}. Comparisons with the naive deconvolution demonstrate that Eq.~(\ref{fil_deconv}) is far more accurate,
especially for filaments (or sources) with shallower slopes $\gamma\lesssim 2$. The inaccuracies of Eq.~(\ref{fil_deconv}) are mostly
within a few percent, except near the lower bounds of resolvedness (see below), where errors become larger. For Gaussian profiles
($\gamma\rightarrow\infty$), the formula is accurate within 12\%.

The concept of resolvedness becomes more complex for ``infinite'' power-law structures \citep{Men'shchikov2023}, because
$\mathcal{R}$ has its lower bound of unity only for very steep (Gaussian) profiles. Indeed, Eq.~(\ref{fil_deconv}) shows that the
lower bound shifts to larger values $\mathcal{R}_{\rm min\!} = B^{1/A}$ for ``infinite'' power-law profiles, depending on $\gamma$
(Fig.~\ref{deconvolinf}). This is because convolution of filaments with shallower slopes spreads their peaks over significantly
wider areas than for steeper profiles. To determine whether an observed filament is resolved, it is necessary to compare its
measured width $H$ to $O B^{1/A}$, not just to the angular resolution $O$. For a formal definition, it makes sense to require that
a resolved power-law filament must have its deconvolution function $\breve{H}H^{-1\!}\gtrsim 0.7$, which is equivalent to the condition
$H\gtrsim 2OB^{1/A}$ or $\mathcal{R}\gtrsim 2B^{1/A}$. For example, Fig.~\ref{deconvolinf} demonstrates that the filament with $\gamma =
0.5$ is unresolved for $\mathcal{R}\lesssim 3.5$ and the flatter one with $\gamma = 0.125$ is unresolved for $\mathcal{R}\lesssim 10$.

\subsection{``Finite'' (arbitrarily extended) filaments}
\label{finite}

Deconvolution functions for deriving the half-maximum widths $\breve{H}$ and $h$ of filaments with finite footprints of any width
are more complicated than in Eq.~(\ref{fil_deconv}), because these functions now depend on two parameters: $\beta$ and $\xi\equiv
R/h$. We investigated the problem numerically by creating surface density images of filaments from
Eqs.~(\ref{surface_density_fittingfun})--(\ref{hHformula}) parameterized by $h_{\rm T}=$ 44.2{\arcsec}, $\beta_{\rm T}=$ \{0.5, 1,
1.5, 2\}, and $\xi_{\rm T}=$ \{1, 2, 4, 8, 16, 32, 64\}. The model images were convolved with Gaussian beams of half-maximum widths
$O$ ranging from 9.5 to 1730{\arcsec}. We then measured the resulting filament widths $H$ and evaluated the deconvolution functions
$\breve{H}H^{-1}$ and $hH^{-1}$ as functions of the resolvedness $\mathcal{R}$ and the true values of the parameters $\beta$ and
$\xi$.

\begin{figure}
\centerline{\resizebox{0.8\hsize}{!}{\includegraphics{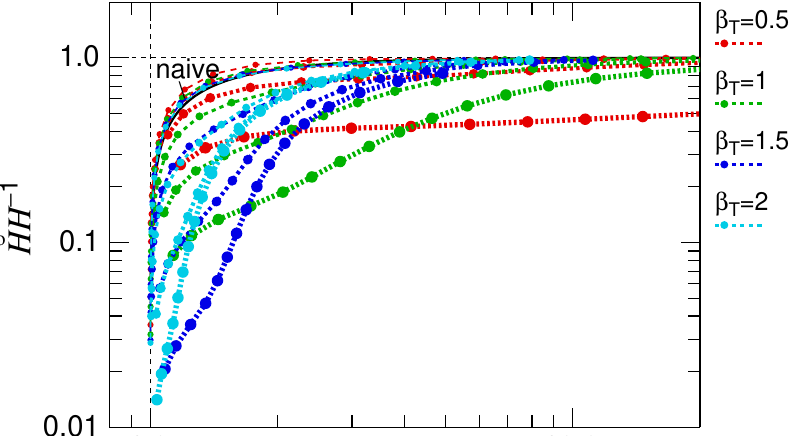}}}
\centerline{\resizebox{0.8\hsize}{!}{\includegraphics{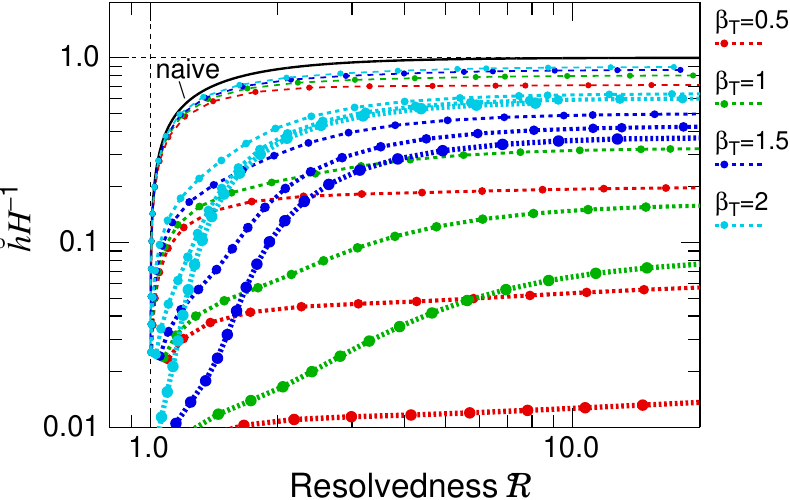}}}
\caption{
Deconvolution functions $\breve{H}H^{-1}$ (\emph{top}) and $\breve{h}H^{-1}$ (\emph{bottom}) for ``finite'' power-law filaments
(Sect.~\ref{finite}). The filled circles connected by curves represent the functions for deriving the deconvolved widths
$\breve{H}$ and $\breve{h}$ from measured widths $H$ of surface density profiles as functions of resolvedness $\mathcal{R}$. Four
styles of dashed/dotted curves of the same color and increasing widths correspond to true extents $\xi_{\rm T}= $ \{1, 4, 16, 64\}.
For reference, the black solid curves show the deconvolution function from Eq.~(\ref{gauss_deconv}). The dashed lines indicate
limiting values on both axes.
} 
\label{deconvolfin}
\end{figure}

\begin{figure*}[ht!]
\centerline{\resizebox{0.3225\hsize}{!}{\includegraphics{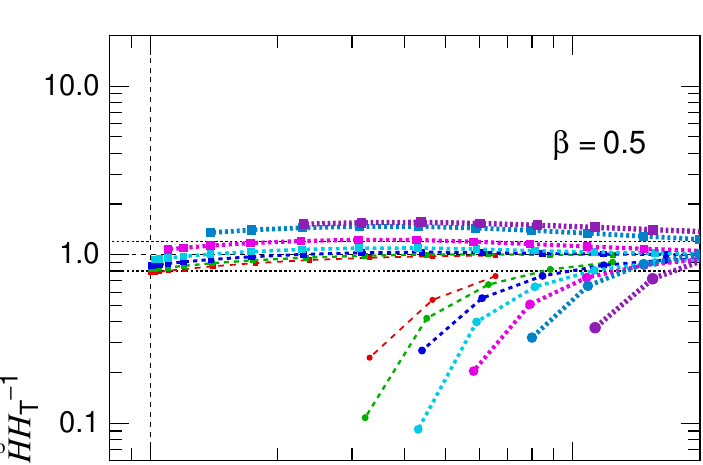}}\!
            \resizebox{0.3081\hsize}{!}{\includegraphics{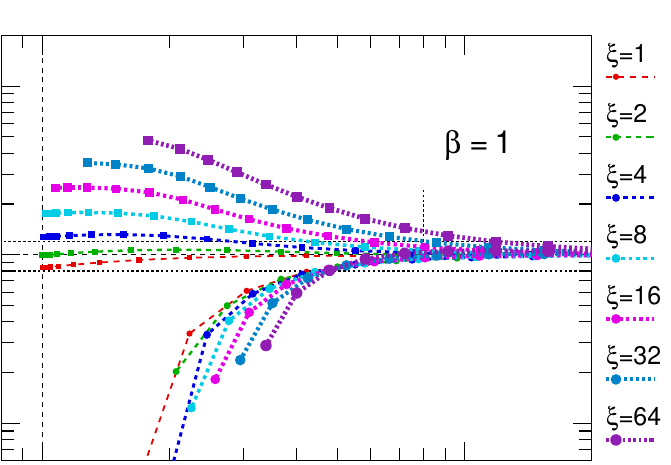}}}
\centerline{\resizebox{0.3225\hsize}{!}{\includegraphics{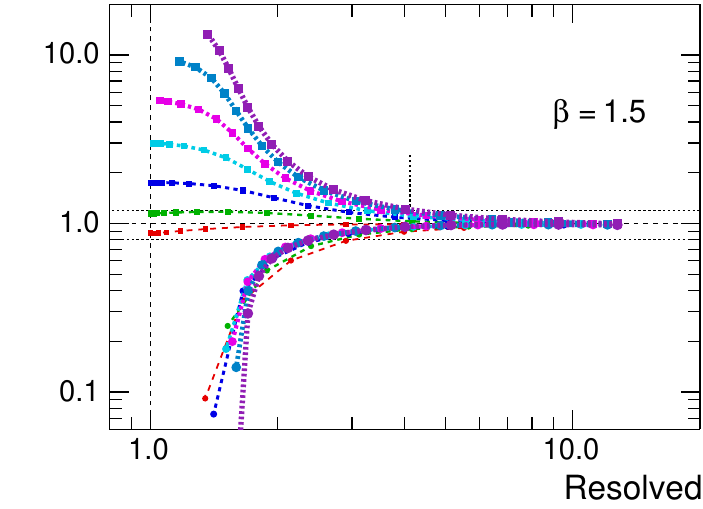}}\!
            \resizebox{0.3081\hsize}{!}{\includegraphics{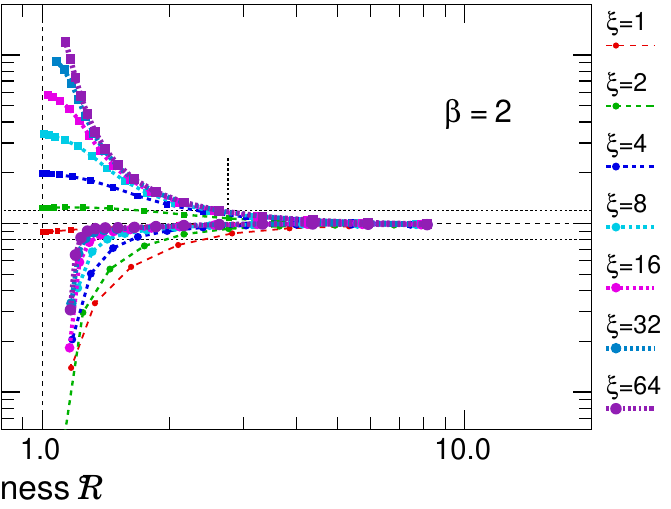}}}
\caption{
Accuracies of the deconvolution functions from Eqs.~(\ref{gauss_deconv}) and (\ref{fil_deconv}) when applied to convolved models of
``finite'' power-law filaments (Sect.~\ref{finite}). Filled squares and circles show the accuracy ratios $\breve{H}H^{-1}_{\rm T}$
for the ``naive'' (Gaussian) and ``infinite'' (filament-specific) deconvolutions, respectively, as functions of resolvedness
$\mathcal{R}$. The seven symbol styles and connecting dashed/dotted curves of the same color with increasing line widths correspond
to extents $\xi_{\rm T}= $ \{1, 2, 4, 8, 16, 32, 64\}. The horizontal dashed line indicates perfect accuracy, and the two
horizontal dotted lines show $\pm 20$\% errors. Vertical dotted lines mark the minimum resolvedness $\mathcal{R}_{20}\approx
1+7\beta^{-2}$ required for reasonably accurate ``naive'' deconvolution of finite power-law filaments.
} 
\label{acc_naive_infinite}
\end{figure*}

The numerical deconvolution functions shown in Fig.~\ref{deconvolfin} display large differences from those in
Fig.~\ref{deconvolinf}. As expected, for the smallest extent $\xi_{\rm T}=1$, they appear roughly similar to that of
Eq.~(\ref{gauss_deconv}) and show only slight dependence on $\beta$. For increasing extents $\xi_{\rm T}$ at shallow slopes
($\beta_{\rm T}=0.5$), the deconvolved widths $\breve{h}$ become much narrower than the measured width $H$, whereas $\breve{h}$ and
$H$ become increasingly similar to each other for steeper slopes. For $\mathcal{R}\gtrsim 4$, the behavior of $\breve{h}$ and
$\breve{H}$ is fully consistent with the relationships between $h$ and $H$ in Fig.~\ref{relationships}. In contrast with the case
of ``infinite'' filaments (Fig.~\ref{deconvolinf}), the lower bound of resolvedness remains $\mathcal{R}_{\rm min}=1$ with no
dependence on the slopes. This is because at sufficiently low angular resolutions, the convolution beam becomes wider than the
entire filament footprint ($O>2R$), and the filament becomes completely unresolved ($H\approx O$). 

As expected, the ``finite'' deconvolution function $\breve{H}H^{-1}$ in Fig.~\ref{deconvolfin} becomes progressively more similar
to the ``infinite'' one shown in Fig.~\ref{deconvolinf} for profiles with larger extents ($\xi_{\rm T}\gtrsim 64$). However, the
practically most important deconvolution function $\breve{h}H^{-1}$ exhibits strong and complicated dependencies on both $\beta$
and $\xi$. The large divergence between $\breve{h}H^{-1}$ and $\breve{H}H^{-1}$ for shallow slopes and large extents
directly reflects the $hH^{-1}$ ratio shown in Fig.~\ref{relationships}: for extended filaments with $\beta\lesssim 2$ and large
$\xi$, the volume density width $h$ is much smaller than $H$, so recovering $h$ from $H$ requires a far larger correction
than recovering $\breve{H}$. Given the significant uncertainties in these parameters when analyzing real observations,
accurate derivation of the volume density width $h$ is challenging and requires careful consideration of the applicable
deconvolution method.

\subsection{Practical considerations}
\label{practicalities}

Despite the challenges outlined above, it is still useful to assess which deconvolution approach yields acceptable accuracy under
different conditions. When analyzing observed filaments, it is necessary to quantify the errors in the widths $\breve{H}$
deconvolved with either the ``naive'' (Gaussian) or ``infinite'' (filament-specific) formula. To evaluate these errors, we applied
Eqs.~(\ref{gauss_deconv}) and (\ref{fil_deconv}) using the widths $H$ of the $\Sigma(r)$ profiles from the convolved ``finite''
filaments and the slope $\gamma$ derived from Eq.~(\ref{betaformula}) for the model $\beta_{\rm T}$ and $\xi_{\rm T}$ values.
Results in Fig.~\ref{acc_naive_infinite} show that shallow profiles with $\beta_{\rm T}=0.5$, as well as steeper profiles with
small extents $\xi_{\rm T}=$ \{1, 2\}, can be deconvolved fairly accurately by the simple Eq.~(\ref{gauss_deconv}) because they are
both compact and steep. When the slope $\beta$ becomes steeper and the extent $\xi$ increases, the ``infinite'' deconvolution
yields increasingly better accuracy because the profiles become more similar to those used to derive Eq.~(\ref{fil_deconv}).

In the context of analyzing real observations, the accuracies of the ``infinite'' deconvolution shown in
Fig.~\ref{acc_naive_infinite} assume that the measured surface density parameters $H$ and $\gamma(\beta,\xi)$ are derived from
fitting the observed profiles with high precision. However, our tests discussed in Sect.~\ref{angular_resolution} demonstrate that
these parameters can have very significant systematic errors when angular resolution is insufficiently high. Therefore, it may be
preferable to apply the ``naive'' deconvolution (which has no free parameters) in the resolvedness range where it provides adequate
accuracy. If we require errors to be within 20\% (a reasonable threshold for many astrophysical applications), then the validity
range for deconvolution with Eq.~(\ref{gauss_deconv}) can be approximated by $\mathcal{R}> \mathcal{R}_{20}\approx 1+7\beta^{-2}$.
However, this relation assumes accurately known $\beta\approx \beta_{\rm T}$. In practice, when $\beta$ itself has large systematic
errors due to limited resolution, this criterion becomes less reliable, and one should exercise caution in interpreting the
deconvolved widths. In such cases, improving the angular resolution of observations is the most robust solution.

\end{appendix}

\end{document}